\documentclass[useAMS,usenatbib]{mn2e}

\usepackage{aecompl}
\usepackage[toc,page]{appendix}

\usepackage{graphicx}
\usepackage{amssymb}
\usepackage{amsmath}
\usepackage{aas_macros}
\usepackage{deluxetable}
\usepackage{lscape}
\usepackage{rotating}
\usepackage[symbol*]{footmisc}



\title{Filamentary fragmentation in a turbulent medium}
\author[S. D. Clarke, A. P. Whitworth, A. Duarte-Cabral and D. A. Hubber]{S. D. Clarke$^{1}$\thanks{E-mail: seamus.clarke@astro.cf.ac.uk }, A. P. Whitworth$^{1}$, A. Duarte-Cabral$^{1}$ and D. A. Hubber$^{2,}$$^{3}$\\$^{1}$School of Physics and Astronomy, Cardiff University, Cardiff, CF24 3AA, UK\\$^{2}$University Observatory Munich, Ludwig-Maximilians-University Munich, Scheinerstr.1, D-81679 Munich, Germany\\$^{3}$Excellence Cluster Universe, Boltzmannstr. 2, D-85748 Garching, Germany}

\newcommand{\OO}{_{_{\rm O}}}
\newcommand{\Su}{_{_{\odot}}}
\begin{document}

\date{}

\pagerange{\pageref{firstpage}--\pageref{lastpage}} \pubyear{2002}

\maketitle

\label{firstpage}

\begin{abstract}

We present the results of smoothed particle hydrodynamic simulations investigating the evolution and fragmentation of filaments that are accreting from a turbulent medium. We show that the presence of turbulence, and the resulting inhomogeneities in the accretion flow, play a significant role in the fragmentation process. Filaments which experience a weakly turbulent accretion flow fragment in a two-tier hierarchical fashion, similar to the fragmentation pattern seen in the Orion Integral Shaped Filament. Increasing the energy in the turbulent velocity field results in more sub-structure within the filaments, and one sees a shift from gravity-dominated fragmentation to turbulence-dominated fragmentation. The sub-structure formed in the filaments is elongated and roughly parallel to the longitudinal axis of the filament, similar to the fibres seen in observations of Taurus, and suggests that the \textit{fray and fragment} scenario is a possible mechanism for the production of fibres. We show that the formation of these fibre-like structures is linked to the vorticity of the velocity field inside the filament and the filament's accretion from an inhomogeneous medium. Moreover, we find that accretion is able to drive and sustain roughly sonic levels of turbulence inside the filaments, but is not able to prevent radial collapse once the filaments become supercritical. However, the supercritical filaments which contain fibre-like structures do not collapse radially, suggesting that fibrous filaments may not necessarily become radially unstable once they reach the critical line-density.   

\end{abstract}

\section{Introduction}%

It has long been known that filaments play a role in the star formation process \citep{SchElm79}. The recent observations by the Herschel Space Observatory have revealed that filaments are prevalent within molecular clouds and that star-forming cores are often embedded within them \citep{And10, Arz13, Kon15, Mar16}. The ubiquity of filamentary structures has lead to numerous papers studying the effects their geometry has on stability and fragmentation \citep{Ost64, Lar85, InuMiy92, Fre14, Cla16}.

Previous work has shown that filaments are prone to fragmentation due to their geometry; small-scale density perturbations have time to collapse locally before global longitudinal collapse occurs \citep{Pon11}. \citet{InuMiy92} have performed an analytic perturbation analysis to investigate how small-scale density perturbations grow in equilibrium filaments. They find that the most unstable perturbation has a wavelength roughly four times the filament's diameter.

However, it is unlikely that a filament forms, stops accreting, relaxes to equilibrium and then proceeds to fragment. Instead, filament and perturbations will co-evolve during a non-equilibrium accretion stage until the filament becomes unstable and fragments. This scenario has been investigated in an earlier work, \citet{Cla16} (hereafter CWH16). Using numerical simulations to investigate the perturbation growth in an accreting filament, CWH16 find that the fastest growing mode is the result of a resonance between radial accretion and longitudinal oscillations. The dominant wavelength is given by

\begin{equation}
\lambda_{_{\rm DOM}} \sim 2 \tau_{_{\rm CRIT}} a\OO,
\label{eq::DOM}
\end{equation}
where $\tau_{_{ \rm CRIT}}$ is the time at which the filament becomes radially unstable and $a\OO$ is the isothermal sound speed.

This fastest growing mode continues to be the dominant mode in simulations seeded with perturbations at multiple wavelengths. Moreover, the fastest growing mode is linked to the environment of the filament, its temperature and its accretion rate. This allows observers to determine the age, $\tau_{_{\rm AGE}}$, and mean accretion rate, $<\dot{\mu}>$, of a filament which is fragmenting in a roughly periodic manner,

\begin{equation}
\tau_{_{\rm AGE}} \geq \tau_{_{\rm CRIT}} \simeq \frac{\lambda_{_{\rm CORE}}}{2 a\OO},
\label{eq::time}
\end{equation}
and
\begin{equation}
<\dot{\mu}> \simeq \frac{2a\OO \mu}{\lambda_{_{\rm CORE}} },
\label{eq::mu}
\end{equation}
where $\lambda_{_{\rm CORE}}$ is the measured core separation and $\mu$ is the filament's line-density.

\citet{TafHac15} have recently observed the star forming filament L1495 in Taurus, and find that it can be decomposed into velocity-coherent sub-filaments, which they term 'fibres'. On the basis of these results, they propose that filaments first fragment into fibres and that these fibres then go on to fragment into cores, a model they call \textit{fray and fragment}. Recent simulations by \citet{Smi14} suggest that fibres do not form from the fragmentation of filaments, but rather form in the surrounding cloud and are then swept up by large scale motions to form a single column density structure. At present it is unclear which mechanism is dominant, or what role fibres play in filamentary fragmentation.

In this paper, we present numerical simulations of initially sub-critical filaments which accrete from a turbulent medium. Unlike the simulations presented in CWH16 in which density perturbations were placed in the accretion flow, the simulations in this paper use a turbulent velocity field to seed density perturbations. In Section \ref{SEC:NUM}, we detail the simulation setup and the initial conditions used; in Section \ref{SEC:RES} we present the results of these simulations; in Section \ref{SEC:DIS}, we discuss the significance of these results and compare to previous work and observations; and in Section \ref{SEC:CON}, we summarise our conclusions. 

\section{Numerical Setup}\label{SEC:NUM}%

The simulations presented in this paper are performed using the Smoothed Particle Hydrodynamics (SPH) code \textsc{Gandalf} (Hubber et al. in prep.). The simulations invoke both self-gravity and hydrodynamics, with an isothermal equation of state. The temperature, $T\OO$, is taken to be $ 40 \, \rm{K}$ throughout so as to compare to the fiducial case presented in CWH16; this results in an isothermal sound speed of $\sim \, 0.37 \, \rm{km/s}$, assuming solar metallicity. Grad-h SPH \citep{PriMon04} is implemented, with $\eta=1.2$, so that a typical particle has $\sim 58$ neighbours. 

Sink particles are implemented as described in \citet{Hub13} using the sink creation density, $\rho_{_{\rm SINK}} = 10^{-15} \, \rm{g \, cm^{-3}}$. However, as well as the standard sink creation criteria described in \citet{Hub13}, an additional criterion is used which mitigates the over-production of sink particles in well-resolved clumpy accretion flows onto an existing sink particle. To understand this additional criterion, consider a region in an accretion flow that is gravitationally bound, and an SPH particle, $i$, in it that satisfies all of the standard sink creation criteria. However, if this region is moving towards an existing sink particle, $j$, it may be accreted by sink $j$ before the region's free-fall time has elapsed. If such a region is replaced with a new sink particle, one would end up with two smaller sink particles where there ought to be only one large sink particle. Thus, a candidate SPH particle's free-fall time must be less than the accretion timescale if it is to be replaced by a sink,
\begin{equation}
\tau_{ff,i} < \frac{|\mathbf{r}_{ij}|}{\mathbf{r}_{ij} \cdot \mathbf{v}_{ij}},
\end{equation} 
\begin{equation}
\rho_{i} > \frac{3 \pi}{32G} \left( \frac{\mathbf{r}_{ij} \cdot \mathbf{v}_{ij}}{|\mathbf{r}_{ij}|}\right)^2.
\end{equation}
Our simulations have been tested with sink creation densities ranging from $10^{-12} \, \rm{g \, cm^{-3}}$ to $10^{-17} \, \rm{g \, cm^{-3}}$, and we are confident that a sink creation density of $10^{-15} \, \rm{g \, cm^{-3}}$ is well converged \footnote[1]{For an overview of the standard sink creation criteria and a convergence test showing the implementation of the new criteria see appendix \ref{APP:SINK}.}.

The computational domain is open in the $x$- and $y$- directions, but periodic in $z$, the long axis of the filament (see W{\"u}nsch et al. in preparation, for the implementation of the modified Ewald field). This allows us to study the fragmentation of an infinitely long filament, and ignore the complicating effects of global longitudinal collapse \citep{ClaWhi15}.

To generate the initial conditions, a cylindrical settled glass of particles with uniform density is stretched so that it reproduces the density profile, 
\begin{equation}
\rho(R,z) \, = \, \frac{\rho\OO R\OO}{R}.
\label{eq::rho}
\end{equation}  
Where $R \equiv (x^2 + y^2)^{1/2}$ and $\rho\OO$ is the density at $R\OO = 0.1 \rm pc$. Note that this is a uniform density profile in the longitudinal $z$ direction. The initial velocity field is
\begin{equation}
\mathbf{v} = - v\OO \mathbf{\hat{R}} + \mathbf{v}_{_{\rm TURB}}.
\label{eq::vel} 
\end{equation}
The first term of the velocity field is identical to the field used in CWH16, a radially convergent velocity field to model the formation of a filament. The second term is a turbulent velocity field which is added to the gas to seed density perturbations. The characteristics of this turbulent field are discussed below in section \ref{SSEC:TURB}. 

Values of $\rho\OO = 150 \, \rm{M\Su \, pc^{-3} }$ and $v\OO = 0.75 \, \rm{ km/s}$ are used, as these are the fiducial values in CWH16; this corresponds to a mass accretion rate onto the filament of $\sim 70 \, \rm{M\Su \, pc^{-1} \, Myr^{-1}}$, consistent with accretion rates inferred observationally \citep{Pal13, Kir13}.

The turbulent cylinder of particles has a radius of $r_{_{\rm MAX}} = 1 \, \rm pc$ and a length of $ L = 3 \, \rm pc$. This provides a sufficiently large computational domain to allows us to study a wide range of possible perturbation wavelengths, while maintaining good resolution.

Artificial viscosity is needed to capture shocks, but it is known to overly dampen motions on the order of the smoothing length. Therefore, the simulations use the time-dependent artificial viscosity described in \citet{MorMon97} to lessen this effect.

After the first sink particle forms, the simulations run until the sinks have accreted roughly $10\%$ of the initial gas mass. After this time, feedback mechanisms such as protostellar jets are expected to play an important role, and, as these processes are not modelled, the simulations are terminated. 

The simulations are run with 450,000 particles; each SPH particle has a mass of $6.3 \times 10^{-4} \, \rm{M\Su}$, resulting in a conservative mass resolution of $6.3 \times 10^{-2} \, \rm{M\Su}$. Tests with different numbers of particles show that the simulations are converged\footnote[2]{The results of these simulations are presented in appendix \ref{APP:RES}.}.

\subsection{Turbulent velocity fields}\label{SSEC:TURB}%

The initial turbulent velocity field, $\mathbf{v}_{_{\rm TURB}}$, is generated using the procedure described in \citet{Lom15}. The turbulent velocity field is defined by a power spectrum, $P(k) \propto k^{-\alpha}$, where $k$ is the velocity mode's wavenumber. Burgers turbulence, $\alpha = 4$, is appropriate for highly supersonic turbulence, while $\alpha = 11/3$, known as Kolmogorov turbulence, is appropriate for subsonic turbulence in an incompressible fluid (see \citet{Fed13} for a discussion of the difference between Burgers and Kolmogorov spectra.) 

For a compressible fluid with transonic turbulence, the exponent is expected to lie between these two values. In both the Burgers and Kolmogorov cases, the majority of the power lies in the small wavenumber, large wavelength modes and so the exact choice of exponent is not very significant. This is shown to be the case when the exponent is greater than 3 in prestellar core simulations presented by \citet{Wal12}. For this work we use $\alpha = 4$.

We do not include the longest wavelength modes in the $x$ and $y$ directions, $\mathbf{k} = (1,0,0)$ and $(0,1,0)$, as the first term of equation \ref{eq::vel} can be considered as an imposed velocity mode in those directions.  

We consider the effect that the ratio of compressive to solenoidal kinetic energy has on filament fragmentation. Purely compressive turbulence is defined as a curl-free velocity field and purely solenoidal turbulence is defined as a divergence-free velocity field. The distinction between these types of turbulent fields, and their effects on star formation, has been explored in many papers \citep[e.g.][]{Fed08,Fed10a,FedKle12}.

A velocity field with a user-defined ratio of compressive to solenoidal kinetic energy can be constructed using Helmholtz's theorem (see \S2.1.3 of \citet{Lom15} for more details). We consider two cases, purely compressive and a `natural' mix, i.e. where the solenoidal modes have twice the power of the compressive modes. In the nomencluture used by \citet{Lom15} these two cases correspond to $\delta_{\rm sol} = 0$ and $\delta_{\rm sol} = 2/3$ respectively.  

A turbulent field is defined by its power spectrum, which determines the distribution of energy as a function of wavenumber, and the three dimensional velocity dispersion, $\sigma_{\rm 3D}$, which determines the total amount of energy in the velocity field. We consider three cases, $\sigma_{\rm 3D} = 0.1, \, 0.4, \, 1.0 \, \rm{km/s}$, which correspond to Mach numbers of $\mathcal{M} \sim 0.25, \, 1, \, 2.5$ respectively. 

Each turbulent field is generated using a random number seed, and we have performed 10 realisations of each pair of parameters, $\delta_{\rm sol}$ and $\sigma_{\rm 3D}$, resulting in a total of 60 simulations.

\section{Results}\label{SEC:RES}%

\begin{figure}
\centering
\includegraphics[width = 0.98\linewidth]{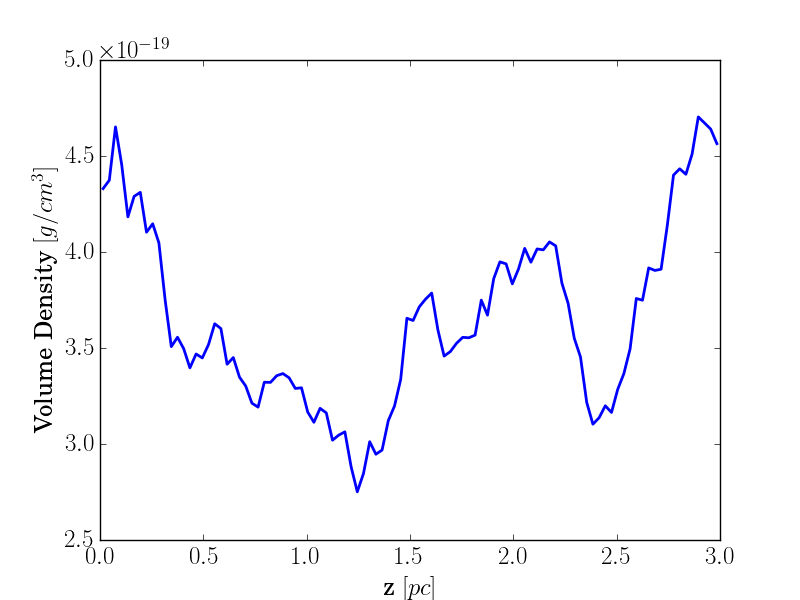}
\caption{The volume density profile along the $z$-axis at $t=0.4 \, \rm Myr$ showing density perturbations on many different scales. In this case, the turbulent field has a natural mix of compressive and solenoidal modes, $\delta_{\rm sol} = 2/3$, and a subsonic velocity dispersion, $\sigma_{\rm 3D} = 0.1 \, \rm{km/s}$.}
\label{fig::denpro}
\end{figure}

The turbulent field produces density perturbations on many different length scales within the filament (for one example, see figure \ref{fig::denpro}). In all simulations, the filament is sub-critical during early times and has a typical width of $\sim 0.1 \, \rm pc$. Gas continues to accrete onto the filament, until it reaches the critical line density and the regions of high density, formed due to the turbulent velocity field, collapse to form sink particles. 

\begin{figure}
\centering
\includegraphics[width = 0.98\linewidth]{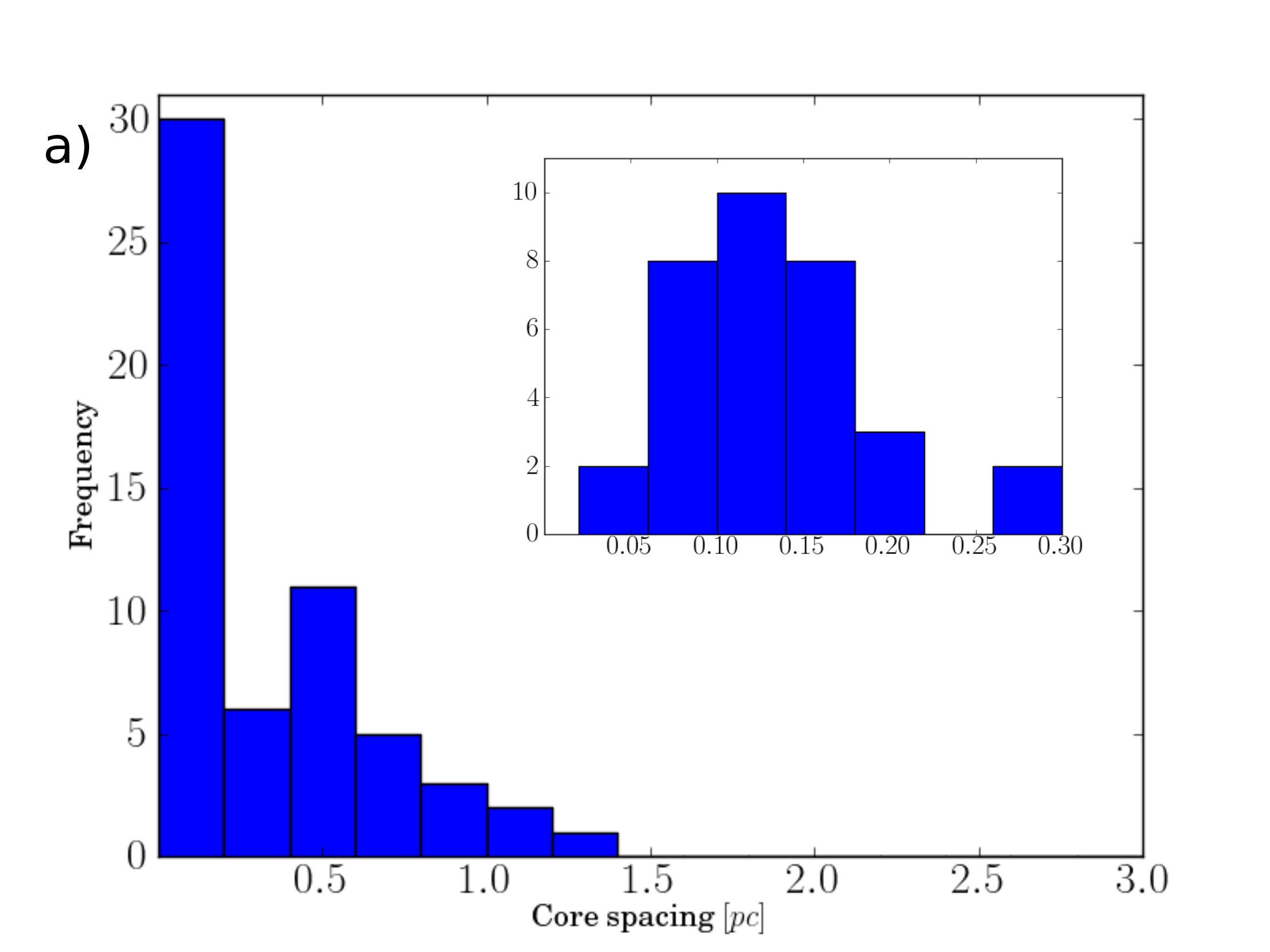}
\includegraphics[width = 0.98\linewidth]{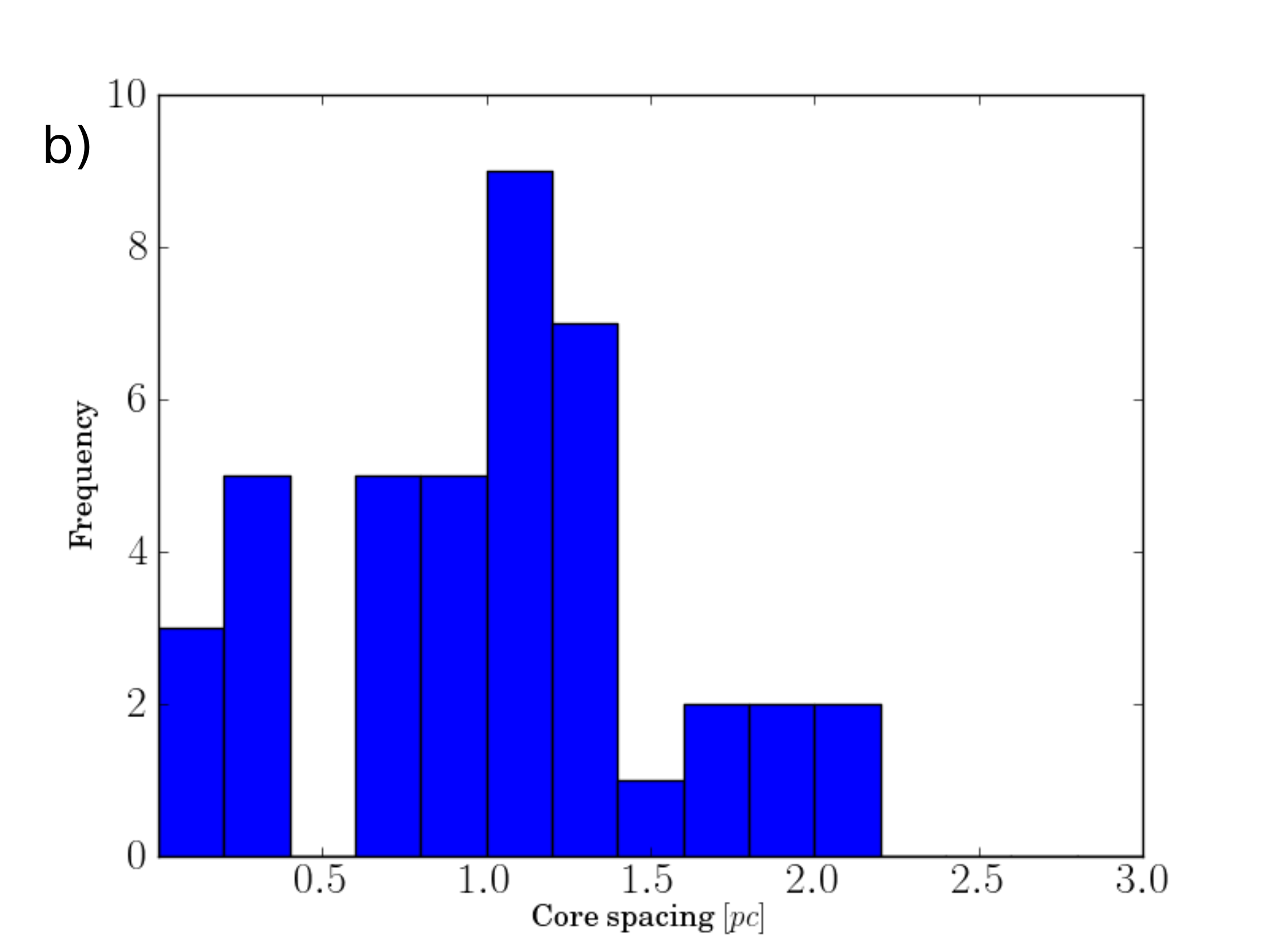}
\includegraphics[width = 0.98\linewidth]{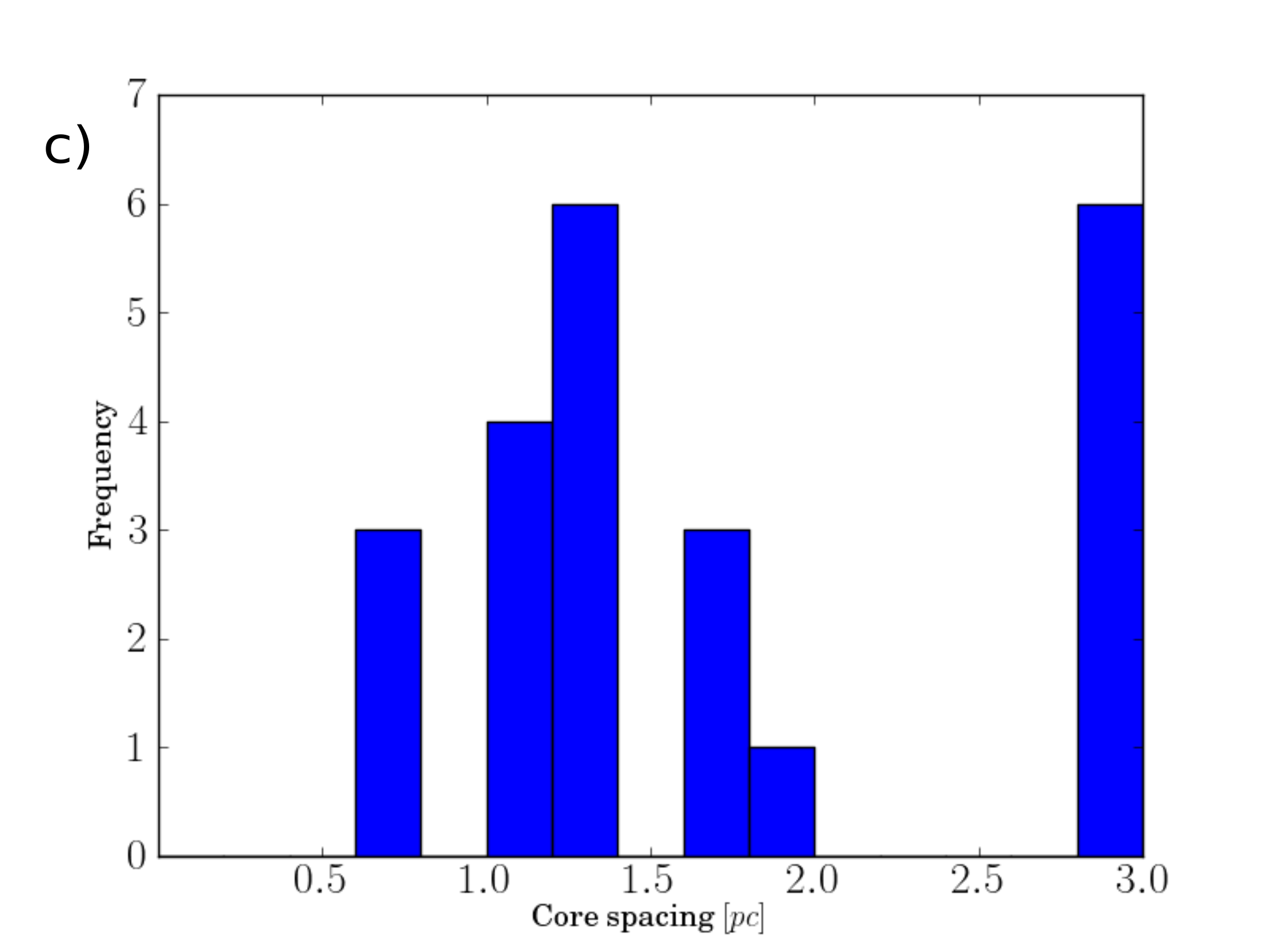}
\caption{Histograms showing the distribution of core separation distances for the set of simulations initialized with a natural mix of turbulence, $\delta_{\rm sol} = 2/3$, for (a) $\sigma_{\rm 3D} = 0.1 \, \rm{km/s}$, (b) $\sigma_{\rm 3D} = 0.4 \, \rm{km/s}$ and (c) $\sigma_{\rm 3D} = 1 \, \rm{km/s}$. Figures a, b and c have 58, 41 and 23 separations respectively. The sub-sonic case (a) has an inset to show the distribution of small scale separations ($<$ 0.3 pc) in more detail. Note that a separation of 3 $\rm pc$ is recorded when the filament has only produced one core, since this is the length of the computational domain and hence the length of the filament.}
\label{fig::disnat}
\end{figure}

\begin{figure}
\centering
\includegraphics[width = 0.98\linewidth]{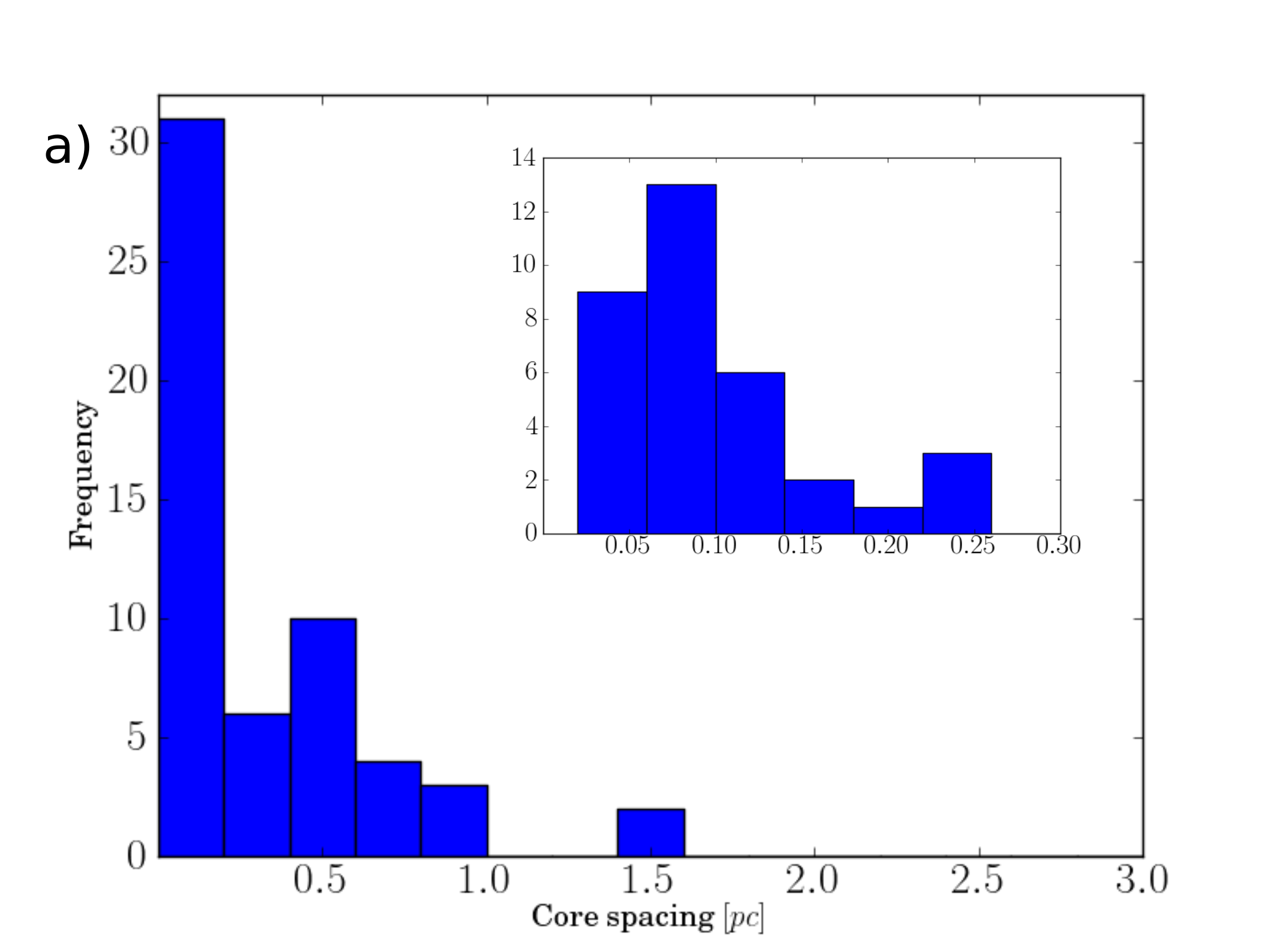}
\includegraphics[width = 0.98\linewidth]{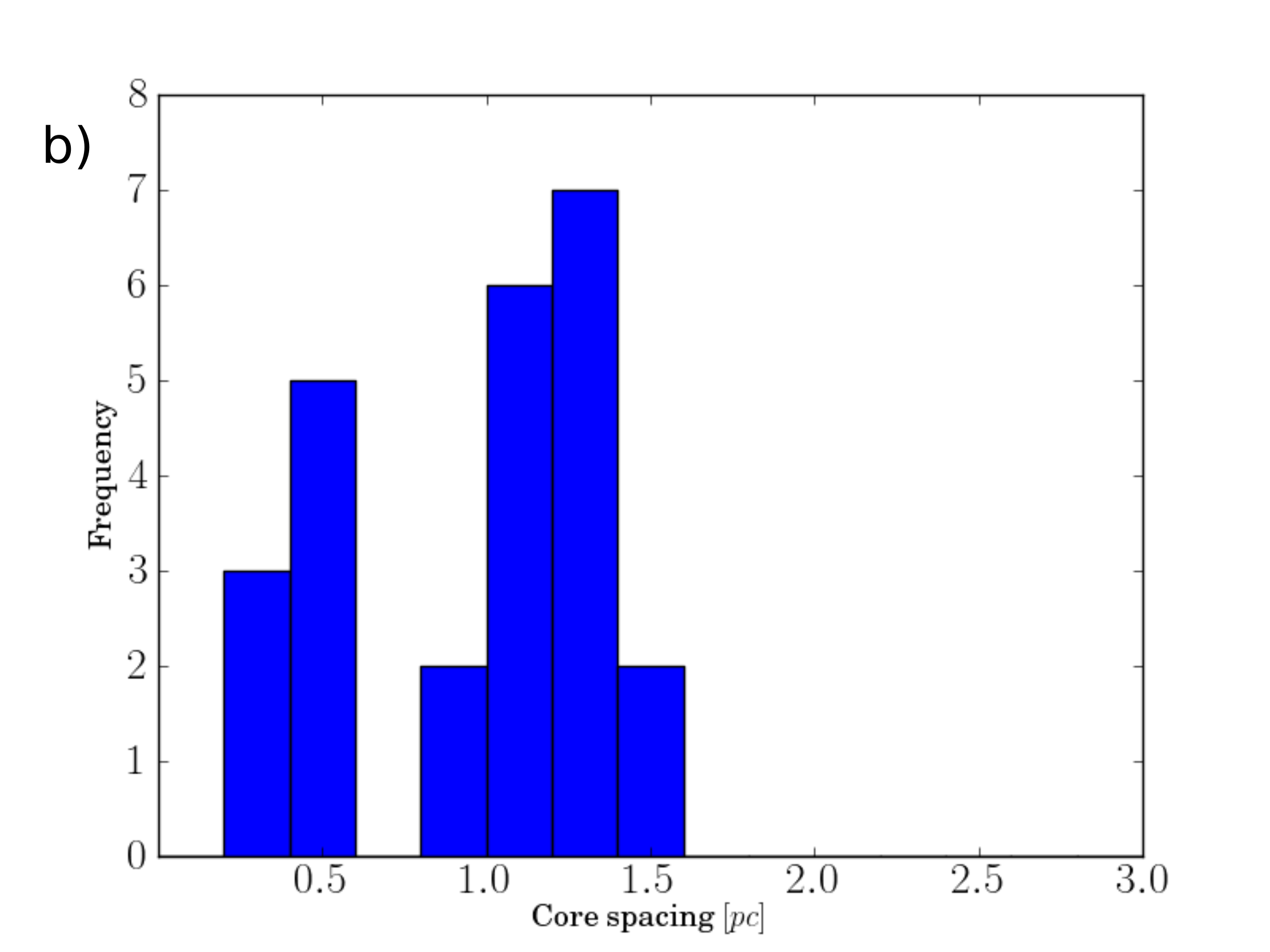}
\includegraphics[width = 0.98\linewidth]{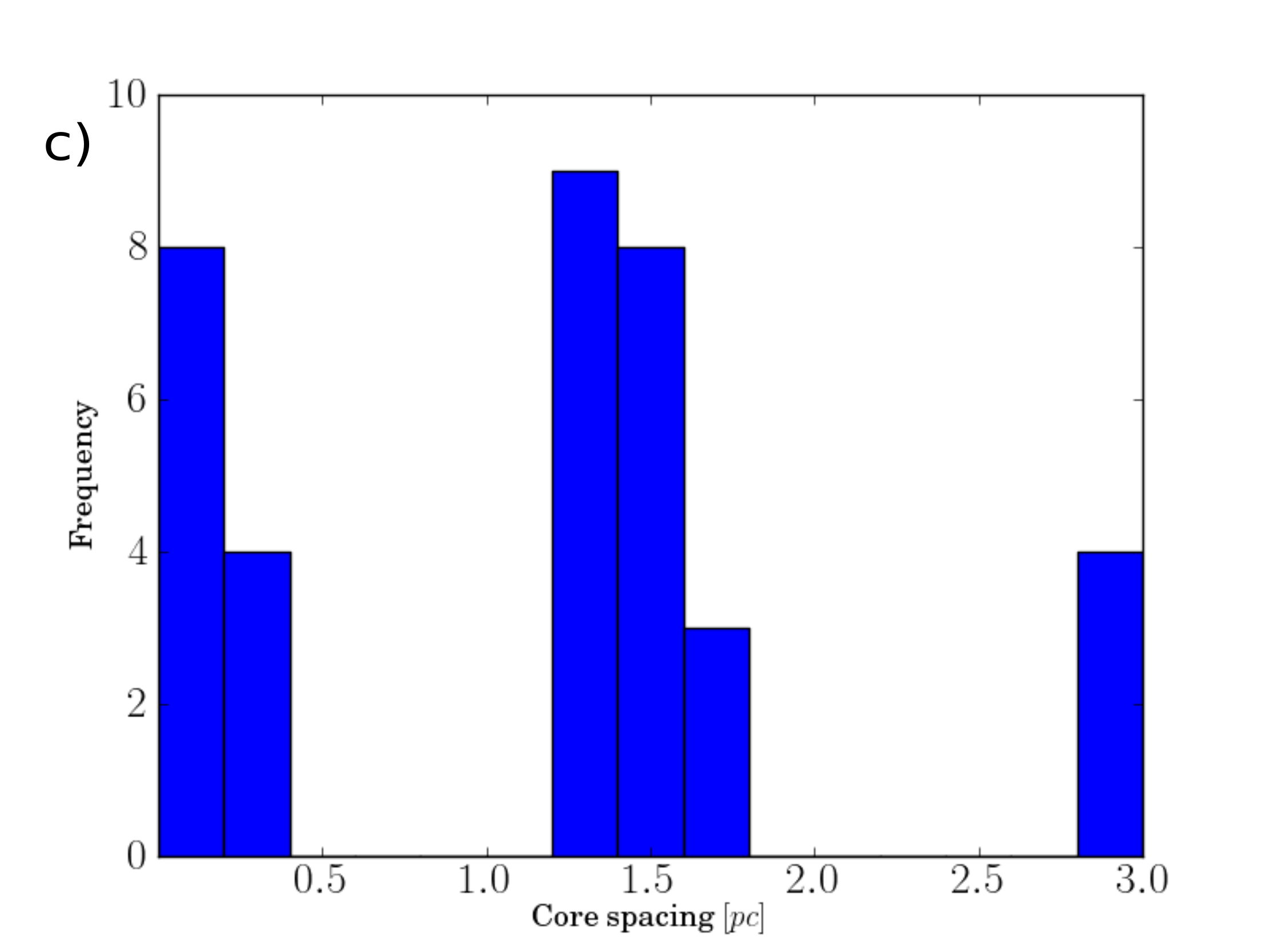}
\caption{Histograms showing the distribution of core separation distances for the set of simulations initialized with purely compressive turbulence, $\delta_{\rm sol} = 0$, for (a) $\sigma_{\rm 3D} = 0.1 \, \rm{km/s}$, (b) $\sigma_{\rm 3D} = 0.4 \, \rm{km/s}$ and (c) $\sigma_{\rm 3D} = 1 \, \rm{km/s}$. Figures a, b and c have 56, 25 and 36 separations respectively. The sub-sonic case (a) has an inset to show the distribution of small scale separations ($<$ 0.3 pc) in more detail. Note that a separation of 3 $\rm pc$ is recorded when the filament has only produced one core, since this is the length of the computational domain and hence the length of the filament.}
\label{fig::discomp}
\end{figure}

To study the fragmentation of a filament we work out the sink separation at the end of the simulation when 10$\%$ of the mass is in sink particles. For each pair of parameters, $\delta_{\rm sol}$ and $\sigma_{\rm 3D}$, we work out the sink separations for all 10 realisations to produce a separation distribution. All separations which are below 0.02 pc are removed; this is because we wish to study filament fragmentation into cores, and separations below $\sim 4000$ AU are clearly due to core fragmentation resulting in multiple systems. Figure \ref{fig::disnat} shows the histograms of the separation distributions for the natural-mix turbulence, $\delta_{\rm sol} = 2/3$, and figure \ref{fig::discomp} shows the same information for the purely compressive turbulence, $\delta_{\rm sol} = 0$. The information from each of the 60 simulations is summarised in a set of tables found in appendix \ref{APP:TAB}.

All 6 histograms seen in figures \ref{fig::disnat} and \ref{fig::discomp} show a more complicated distribution than that found in \citet{Cla16} (their figure 8); turbulence has thus altered the fragmentation of the filaments considerably.

For the subsonic case (figures \ref{fig::disnat}a and \ref{fig::discomp}a), the distributions from the natural mix and purely compressive turbulence share the same morphology. Both core separation distributions show two peaks at the same locations, one at $\sim 0.1 \, \rm pc$ and the other at $\sim 0.5 \, \rm pc$. The small insets show the small scale separations, $< 0.3 \, \rm pc$, in more detail for both values of $ \delta_{\rm sol}$. Here one can see that there is a small difference in the two distributions, the natural mix produces a peak just above 0.1 pc, while the purely compressive turbulence produces a peak just below 0.1 pc.

The transonic case (figures \ref{fig::disnat}b and \ref{fig::discomp}b) is more complicated. The natural mix turbulence has a wide range of core separations and a peak at $\sim 1 \, \rm pc$. The purely compressive turbulence produces a bimodal distribution with peaks at $\sim 0.5 \, \rm pc$ and $\sim 1.2 \, \rm pc$. 

The supersonic case (figures \ref{fig::disnat}c and \ref{fig::discomp}c) produces two very different distributions. The natural mix turbulence causes a number of filaments to only form one core (the large peak in core separations at 3 pc), those that do fragment into numerous cores do so with a wide range of separation distances with a peak $\sim 1.2 \, \rm pc$. More filaments fragment into multiple cores when the turbulence is purely compressive, and those that do, do so with a bimodal distribution with peaks at $\sim 1.5 \, \rm pc$ and $\sim 0.1 \, \rm pc$.

\section{Discussion}\label{SEC:DIS}%

The nature of the turbulence plays an important role in the fragmentation of a filament when the turbulence is transonic or supersonic. Compressive turbulence produces narrower and more clearly defined core separation distributions, and so is more likely to lead to quasi-periodically fragmented filaments. Natural turbulence produces wider distributions with less well defined peaks. 

\begin{figure}
\centering
\includegraphics[width = 0.98\linewidth]{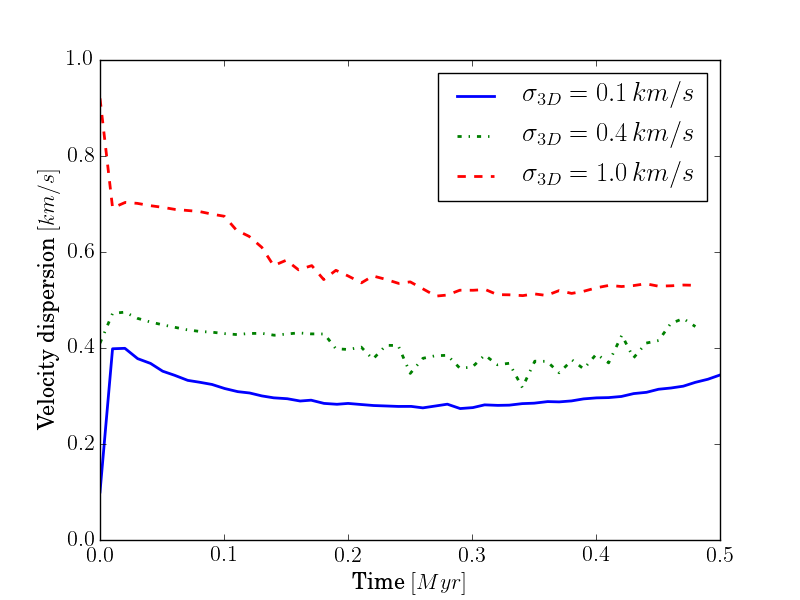}
\caption{Average velocity dispersion as a function of time for the subsonic (solid blue), transonic (dot-dashed green) and supersonic (dashed red) for natural mix turbulence simulations. In all 3 cases the velocity dispersion tends to stabilise at a roughly transonic level, $\sim 0.4 \pm 0.15 \, \rm{km/s}$.}
\label{fig::sig}
\end{figure} 

\begin{figure}
\centering
\includegraphics[width = 0.98\linewidth]{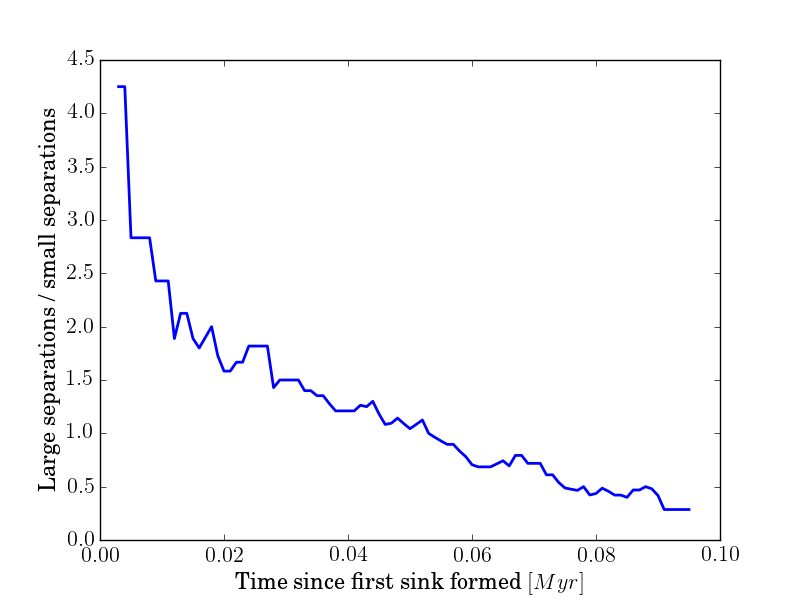}
\caption{Ratio of the number of large scale core separations ($>$ 0.3 pc) over the number of small scale core separations ($<$ 0.3 pc) as a function of time for subsonic natural mix turbulence simulations. Large scale separations dominate at early times, but small scale separations become more common very quickly, on the order of a tenth of a megayear.}
\label{fig::ratio}
\end{figure}

\subsection{Initially subsonic turbulence}

For subsonic turbulence, the nature of the turbulence does not play a significant role in the fragmentation. This can easily be understood by considering the fact that the ratio of initial gravitational potential energy over initial turbulent energy for these simulations is $\sim 100$, thus the dynamics are dominated by gravity. 

Both types of turbulence show bimodal distributions, neither peak corresponding to that suggested by the most unstable mode of a filament, given by equation \ref{eq::DOM}. Taking $a\OO = 0.37 \, \rm{km/s}$ and $\tau_{_{\rm CRIT}} = 0.45 \, \rm Myr$, one would expect to see a peak at $\sim 0.3 \, \rm pc$. These are the values for the fiducial case in CWH16, which share the same accretion rate and temperature as the simulations presented here.

However, equation \ref{eq::DOM} uses the isothermal sound speed which is likely to be inappropriate in a turbulent environment. In the presence of turbulence, one ought to use the effective sound speed defined as $a_{_{\rm EFF}} = \sqrt{a\OO^{2} + \sigma_{1D}^{2} }$, where $\sigma_{1D}$ is the one-dimensional velocity dispersion. For isotropic turbulence, $\sigma_{1D} \, = \, \sigma_{3D} / \sqrt{3}$, where $\sigma_{3D}$ is the three-dimensional velocity dispersion. Changing $a\OO$ for $a_{_{\rm EFF}}$ in equation \ref{eq::DOM}, the most unstable mode in a turbulent filament is given by

\begin{equation}
\lambda_{_{\rm DOM}} \sim 2 \tau_{_{\rm CRIT}} a_{_{\rm EFF}}.
\label{eq::DOM2}
\end{equation} 

To calculate the three-dimensional velocity dispersion within a filament one must first define the edge of the filament. This is done by finding the azimuthally and longitudinally averaged radial profile of the filament and determining the radius at which the density profile blends into the background profile; for these simulations that is the radius where $\rho \, \propto \, r^{-1}$, the initial accretion flow's density profile given by equation \ref{eq::rho}. All SPH particles which lie within this bounding radius are considered to be part of the filament. The three-dimensional velocity dispersion is calculated using velocities in the cylindrical co-ordinate system, $(r,\,\theta,\,z)$; this is done to avoid artificially high velocity dispersions when one uses Cartesian co-ordinates. At each timestep we find the average three-dimensional velocity dispersion of the 10 realisations for each parameter pair. The resulting velocity dispersions for the natural mix turbulence simulations are shown in Figure \ref{fig::sig}. One sees that, no matter what the initial turbulent velocity dispersion is, the three-dimensional velocity dispersion within the filament tends to approximately the same transonic value, $\sim 0.4 \pm 0.15 \, \rm{km/s}$. The velocity dispersion rises after 0.5 Myrs as the filament begins to fragment. This corroborates the results of \citet{Fed16} who also find transonic velocity dispersion in filaments using different initial conditions and a different numerical method.

Taking a value of 0.4 km/s for $\sigma_{3D}$, the effective sound speed in these filaments is $a_{_{\rm EFF}} \sim 0.45 \, \rm{km/s}$. Radial collapse begins at $\tau_{_{\rm CRIT}} \sim 0.5 \, \rm Myr$. Thus the most unstable mode given by equation \ref{eq::DOM2} is $\sim 0.45 \, \rm pc$, consistent with the large scale peak in the core separation distributions for the initially subsonic filaments. 

The small scale peak at $\sim 0.1 \, \rm pc$ is consistent with the effective Jeans length given by $a_{_{\rm EFF}} / \sqrt{G \rho}$, where $\rho$ is the average density in the filament and $G$ is the gravitational constant. The average density in these filaments is of the order $10^{-19} \, - \, 10^{-18} \, \rm{g/cm^{3}}$, which results in a Jeans length of $ 0.05 \, - \, 0.17 \, \rm pc$. 

If the large scale separation is regulated by a variant of the dispersion relation found by CWH16, it may be used to determine the age and mean accretion rate of a filament. One needs to substitute the effective sound speed for the isothermal sound speed in equations \ref{eq::time} and \ref{eq::mu} resulting in,

\begin{equation}
\tau_{_{\rm AGE}} \geq \tau_{_{\rm CRIT}} \simeq \frac{\lambda_{_{\rm CORE}}}{2 a_{_{\rm EFF}}},
\label{eq::time3}
\end{equation}
and
\begin{equation}
<\dot{\mu}> \simeq \frac{2a_{_{\rm EFF}} \mu}{\lambda_{_{\rm CORE}} },
\label{eq::mu2}
\end{equation}

Filaments assembled from gas with subsonic turbulence, or where gravity dominates over turbulence, therefore possess two preferential fragmentation length scales: a large fragmentation length scale which is consistent with a modified version of the filamentary fragmentation model presented in \citet{Cla16}; and a small fragmentation length scale which is consistent with the effective Jeans length in these filaments. 

It is important to note how these fragmentation length scales evolve. Figure \ref{fig::ratio} shows the number of large scale core separations ($>$ 0.3 pc) divided by the number of small scale core separations ($<$ 0.3 pc) as a function of time since the first sink particle forms. Large scale fragmentation proceeds first but is quickly followed by small scale fragmentation such that 0.1 Myrs after fragmentation begins, there are roughly twice as many small length scale core separations as large length scale core separations. The rapid increase in small scale Jeans length separations after collapse begins reduces the periodicity of the fragmentation. The fact that the phase of a filament's life when it is dominated by periodic fragmentation is short-lived may explain why quasi-periodically fragmented filaments are rare, despite both equilibrium and non-equilibrium models suggesting they ought to be standard. 

Thus, initially subsonic filaments, or those where gravity dominates over turbulence, fragment in a two-tier hierarchical fashion, where large scale fragmentation occurs first on a length scale determined by filamentary fragmentation and then those structures proceed to fragment on a smaller scale determined by the effective Jeans length. This type of hierarchical multi-tiered fragmentation has recently been observed in the Orion Integral Shaped Filament \citep{Tei16,Kai16b}.

\subsection{Initially transonic turbulence}

When the turbulence in the accretion flow is subsonic, the separation of fragments is largely determined by gravitational fragmentation -- as described in the preceding subsection. However, once the turbulence in the accretion flow is transonic, the separation of fragments is largely determined by the dominant wavelengths in the turbulence. Specifically, since most of the turbulent energy is in the longer wavelength modes, there is a clustering of separations in the range $1.0 \, \rm pc$ to $1.5 \, \rm pc$, corresponding to the modes with $k_z$ = 2 to 3 (recall that the box is $3 \, \rm pc$ long). This trend is independent of whether the turbulence comprises purely compressive modes, or a natural mix of compressive and solenoidal modes.

As the fragmentation is dominated by the turbulent field, the exact position of the peak in core separations is likely to be the consequence of our choice of $\lambda_{_{\rm MAX}}$, the maximum wavelength in the turbulent field, and is a numerical effect. If $\lambda_{_{\rm MAX}}$ were reduced the position of the peak would likely move to a lower value. This effect has been studied in detail for the case of a turbulent pre-stellar core by \citet{Wal12}.

As well as the `turbulent' peak at $\sim 1.2 \, \rm pc$, the purely compressive turbulence simulations also show a peak at $\sim 0.5 \, \rm pc$, which is consistent with gravitational filamentary fragmentation (equation \ref{eq::DOM2}); this peak is not apparent in the natural-mix turbulence. This suggests that the case of purely compressive turbulence is better described by the gravitational fragmentation model (modified by combining the thermal and turbulent contributions to the velocity dispersion) than the case of natural-mix turbulence, which contains solenoidal modes. The purely compressive case is thus more likely to produce quasi-periodically fragmented filaments.

\begin{figure*}
\centering
\includegraphics[width = 0.45\linewidth]{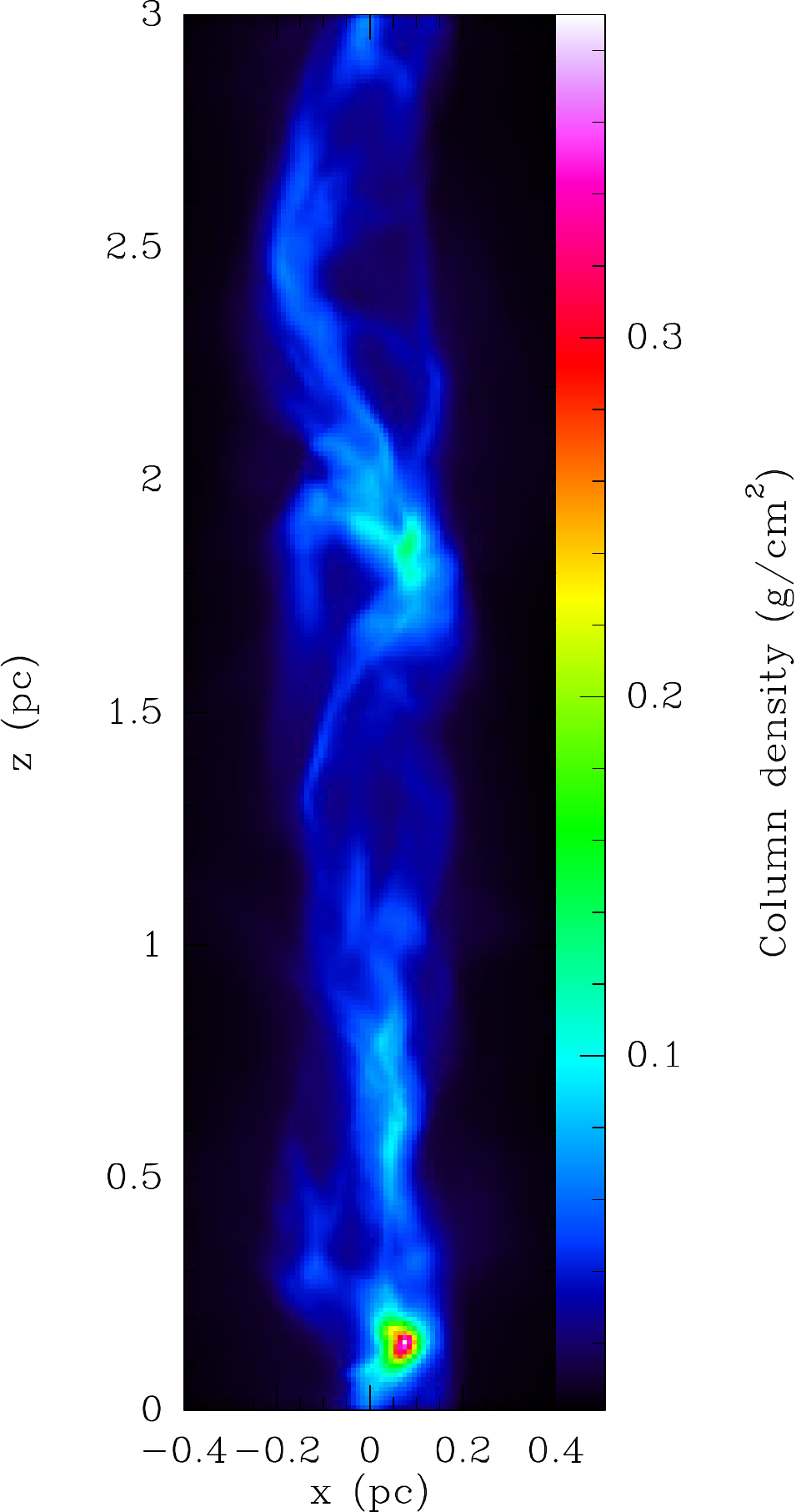}
\hspace{0.25cm}
\includegraphics[width = 0.45\linewidth]{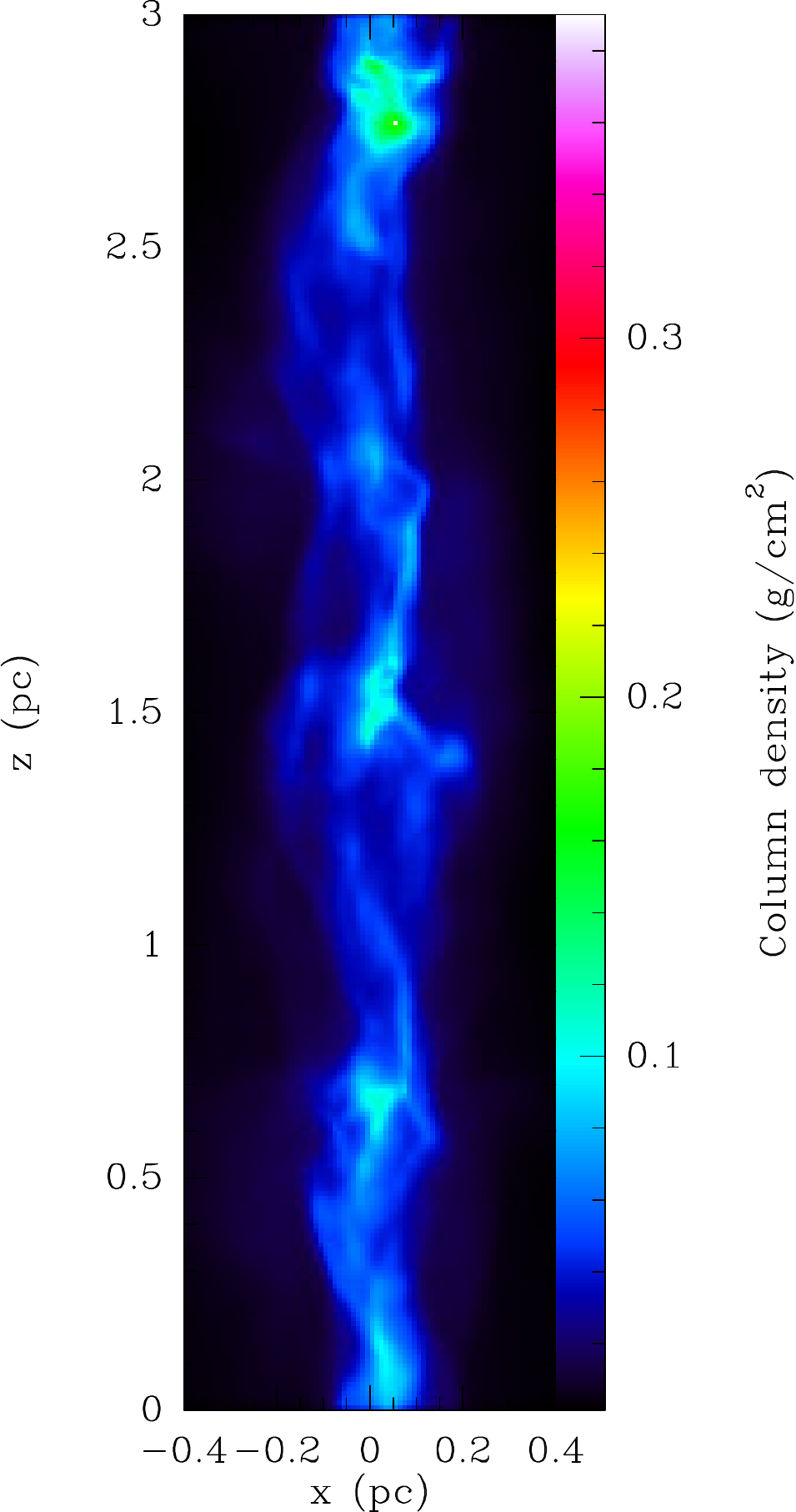}
\caption{Column density maps at $t = 0.5 \, \rm Myr$ for two of the filaments with initially supersonic turbulence, $\sigma_{\rm 3D} = 1 \, \rm{km/s}$ but different types of turbulence. On the left, (a), is the natural mix case, $\delta_{\rm sol} = 2/3$. On the right, (b), is the purely compressive turbulence, $\delta_{\rm sol} = 0$. In both cases we see numerous elongated substructures.}
\label{fig::morph}
\end{figure*}

\subsection{Initially supersonic turbulence}

The initially supersonic cases also show a marked difference due to the nature of the turbulence. This is expected when one considers the fact that the initial turbulent energy is roughly equal to the initial gravitational potential energy, thus the dynamics are markedly influenced by the turbulence. 

Both supersonic cases show the clustering of core separations around $\sim 1.5 \, \rm pc$ due to the turbulent field dominating the fragmentation process. The compressive turbulence produces the narrower of the two distributions, as in the transonic case. There is a peak at very small separations in the purely compressive case which is not apparent in the natural-mix case. This is predominately composed of separations in the range 0.02 - 0.05 pc, i.e. just above the cut-off imposed to remove multiples. It is therefore likely that this peak is due to a number of wide multiples which just exceed the cut-off, and we do not consider these separations further.

A natural mix of modes tends to dampen fragmentation globally, leading to a large number of filaments which only produce one core. Natural-mix turbulence delays the onset of collapse, $\bar{t}_{sink} = 0.717 \, \rm Myr$, more than purely compressive turbulence, $\bar{t}_{sink} = 0.611 \, \rm Myr$. However, the time then taken for $10\%$ of gas to convert into sinks is $0.131 \, \rm Myr$ for the natural mix and $0.128 \, \rm Myr$ for the purely compressive turbulence. Solenoidal modes therefore act as a support on large scales, but are unable to slow star formation once it has started. This is different to the results seen in molecular cloud simulations \citep[e.g.][]{FedKle12} that show that if turbulence is allowed to decay, it strongly reduces the effect of the mixture of turbulent modes.

The increase in initial turbulence also produces a significant change in the morphology of the filaments. This can be seen in column density maps of both the natural mix and the purely compressive cases, which show a number of elongated sub-structures (figure \ref{fig::morph}). The elongated structures seen in the column density maps are real continuous structures in volume density and may be similar to the fibres recently seen in Taurus \citep{Hac13,TafHac15}. A large number of these elongated structures are seen to overlap and merge to form hub-like features, in which most of the cores form; this is somewhat different from the fibres seen in Taurus, which appear to fragment independently of each other.

\begin{figure}
\centering
\includegraphics[width = 0.95\linewidth]{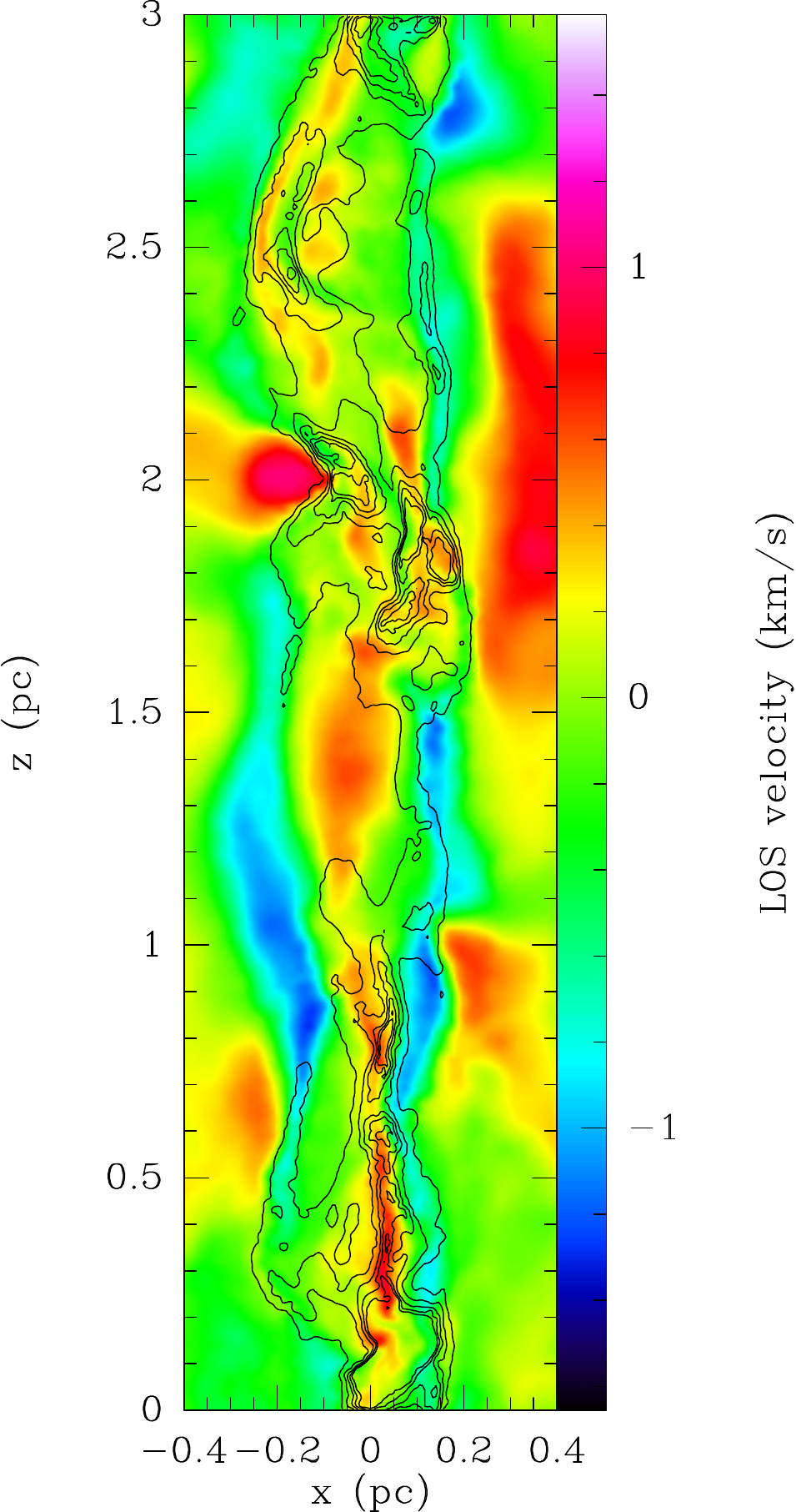}
\caption{A map of the velocity component perpendicular to the plane, the `line-of-sight' velocity $v_y$, at $y=0 \, \rm pc$ with density contours overlaid for an initially supersonic natural mix turbulence simulation at $t=0.5 \, \rm Myr$. The density contours are at $2, 4, 6, 8, 10 \times 10^{-20} \, \rm{g/cm^{3}}$. One can see that the dense elongated structures exhibit smooth and small variations in velocity. Those regions that show large variations are in the diffuse gas, outside the density contours, or in the high density cores which are undergoing collapse.}
\label{fig::los}
\end{figure}

The presence of fibre-like structures may explain why the supersonic cases produce a markedly different core separation distribution from the sub- and transonic simulations, which do not form such structures. In the supersonic cases the fibre-like structures dominate the fragmentation process.

An important point to make is that the fibres identified observationally are found as velocity coherent structures in position-position-velocity (PPV) space while these elongated structures are seen in position-position (PP) space in column density, and in position-position-position (PPP) space in volume density. Without synthetic images it is impossible to make a direct comparison between the fibre-like structures seen in these simulations and the fibres observed. Nonetheless one can consider how the fibre-like structures appear in three-dimensions. Figure \ref{fig::los} shows a slice at $y=0 \, \rm pc$ with the gas velocity in the $y$-direction, the `line-of-sight' velocity, and volume density contours overlaid for one of the natural-mix initially supersonic simulations at $t = 0.5 \, \rm Myr$, the time at which the filament becomes globally thermally supercritical. The density contours are at $2, 4, 6, 8, 10 \times 10^{-20} \, \rm{g/cm^{3}}$. One can see that the velocity in the elongated structures varies smoothly and by small amounts. The regions which exhibit large velocity differences are either outside the contours, and so are diffuse, or associated with compact high density areas which are locally collapsing. Thus, although one cannot be certain that an observer would identify these elongated sub-structures as fibres, they are velocity coherent and continuous in volume-density.

\subsection{Dendrograms}

\begin{figure}
\centering
\includegraphics[width = 0.97\linewidth]{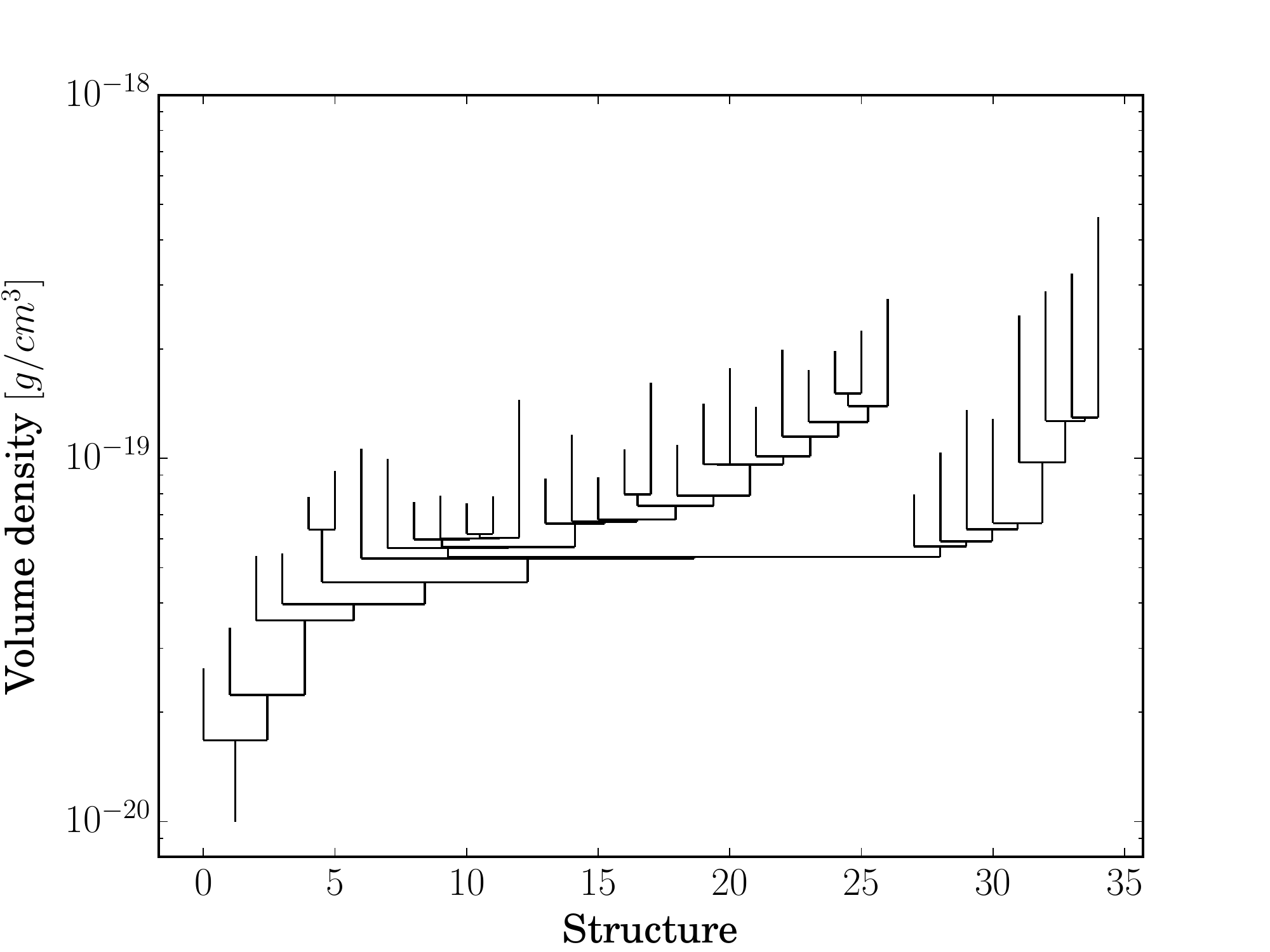}
\includegraphics[width = 0.97\linewidth]{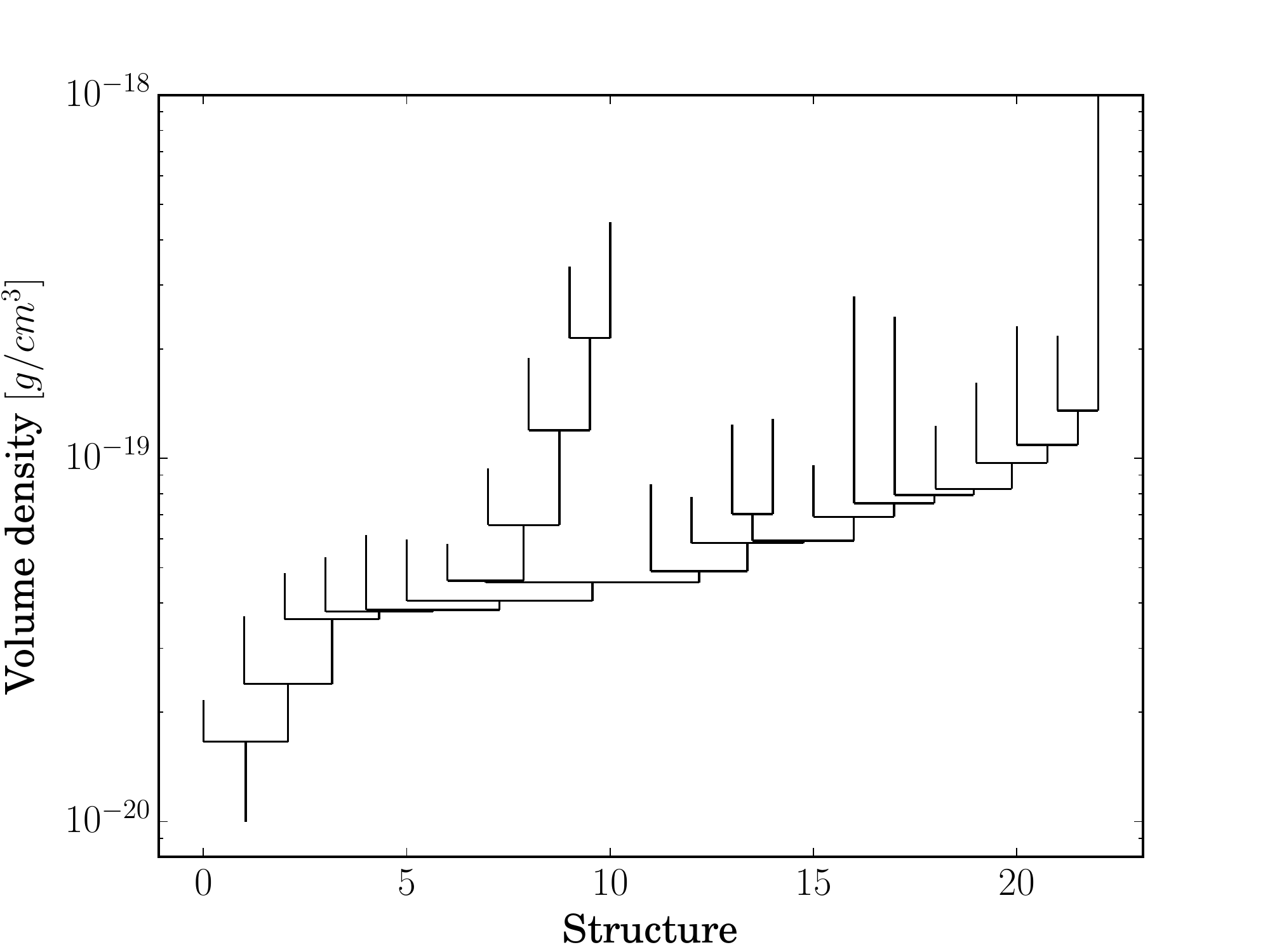}
\caption{Dendrograms constructed using the three-dimensional volume density cubes of (top) a simulation with a natural mix of turbulence, and (bottom) a simulation with purely compressive turbulence. One can see that a natural mix of turbulence produces greater sub-structure, and that sub-structure is more hierarchical, than the simulation with purely compressive turbulence.}
\label{fig::voldendro}
\end{figure}

To investigate quantitatively how hierarchical and sub-structured these filaments are, one may use dendrograms \citep{Ros08}. Here we use the \textsc{python} package \textsc{astrodendro}\footnote[3]{http://www.dendrograms.org}. Dendrograms are constructed from three dimensional volume density cubes with a resolution of 0.01 pc at $t=0.5 \, \rm Myr$, the time at which the filaments become roughly thermally supercritical. The dendrograms are constructed using the parameters: $\rm{min\_value} = 10^{-20} \, \rm {g/cm^{3}}$, the lowest density that a data point must have to be considered when building the dendrogram; $\rm{min\_delta} = 5 \times 10^{-21} \, \rm{g/cm^{3}}$, the minimum density contrast between a leaf and its branch; and $\rm{min\_npix} = 100$, the minimum number of pixels needed for a leaf to be considered a true structure, corresponding to a sphere with a radius $\sim 0.03 \, \rm pc$. Figure \ref{fig::voldendro} shows two example dendrograms, one from a natural mix turbulence simulation and the other from a purely compressive turbulence simulation.

One can obtain two statistics from the dendrograms which quantify the amount of sub-structure and how hierarchical that sub-structure is. The number of leaves in a dendrogram, $N_{_{\rm leaf}}$, tells one how fragmented a structure is, the more leaves the more fragments are present. The number of levels, $N_{_{\rm level}}$, from the highest leaf, that is the leaf which has the largest number of branches between it and the trunk of the dendrogram, tells one how hierarchical the structure is; the more levels, the more nested the sub-structure is. A natural mix of turbulent modes produces both a more fragmented and a more hierarchical structure than the purely compressive turbulence; the mean values and their associated errors: $N_{_{\rm leaf}} = 31.7 \pm 1.6$ and $N_{_{\rm level}} = 20.9 \pm 1.5$ for the natural mix case, and $N_{_{\rm leaf}} = 22.5 \pm 1.8$ and $N_{_{\rm level}} = 14.3 \pm 1.1$ for the purely compressive case. Though the exact values of $N_{_{\rm leaf}}$ and $N_{_{\rm level}}$ change when different values for $\rm{min\_value}$, $\rm{min\_delta}$ and $\rm{min\_npix}$ are used to construct the dendrograms, the result that the natural mix turbulence is more fragmented and hierarchical does not change. One can see this difference in sub-structure in figure \ref{fig::voldendro}.  

\begin{figure}
\centering
\includegraphics[width = 0.97\linewidth]{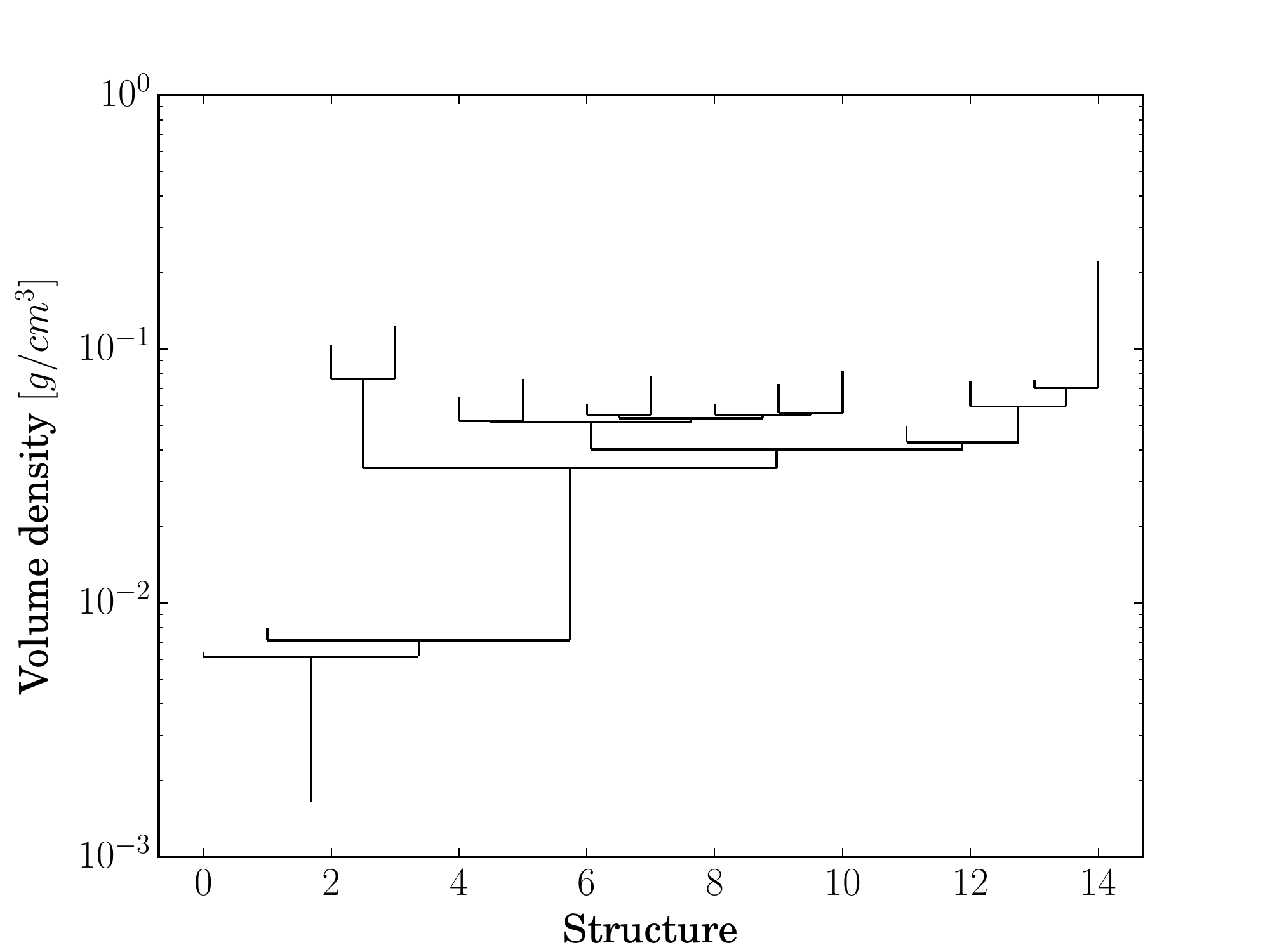}
\includegraphics[width = 0.97\linewidth]{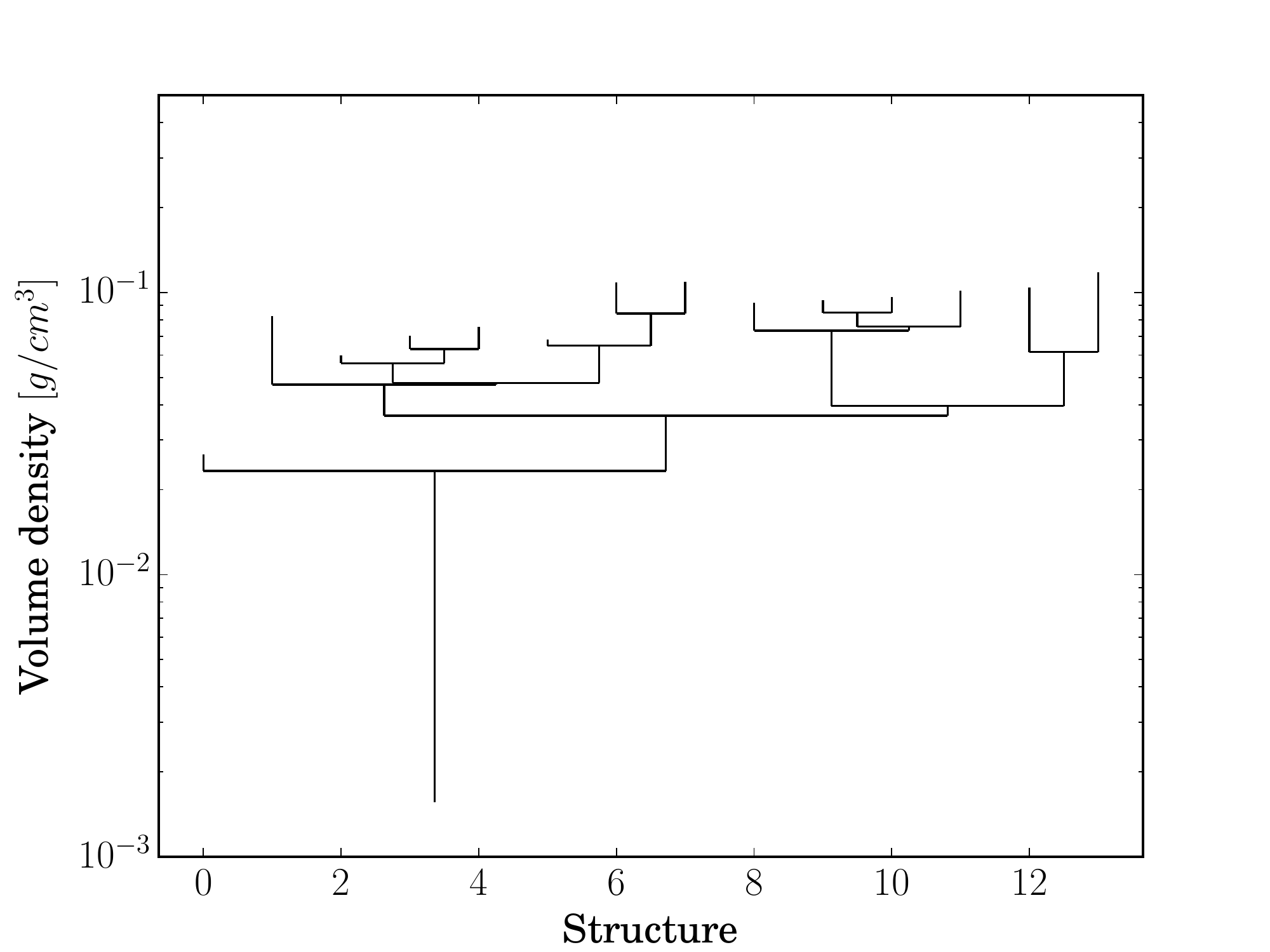}
\caption{Dendrograms constructed using the two-dimensional column density maps of (top) a simulation with a natural mix of turbulence, and (bottom) a simulation with purely compressive turbulence. One can see that there is not a significant difference in either the amount of sub-structure, or how hierarchical the sub-structure is.}
\label{fig::coldendro}
\end{figure}

As an observer does not have access to the three-dimensional volume density structure of a filament, but instead must use the two-dimensional column density, we produce dendrograms using the column density of the simulations. The dendrograms are constructed using: $\rm{min\_value} = 3 \times 10^{-4} \, \rm {g/cm^{2}}$, $\rm{min\_delta} = 1.5 \times 10^{-4} \, \rm{g/cm^{2}}$ and $\rm{min\_npix} = 21$. These values for $\rm{min\_value}$ and $\rm{min\_delta}$ are chosen as they are the choices we made for the three-dimensional volume density dendrograms multiplied by 0.01 pc, the grid resolution. The choice of $\rm{min\_npix}$ was made as it is $100^{2/3}$, the two-dimensional equivalent of our three-dimensional choice. The two cases, natural-mix turbulence and purely compressive turbulence, are indistinguishable from each other when one uses their column density dendrograms: $N_{_{\rm leaf}} = 15.8 \pm 0.7$ and $N_{_{\rm level}} = 9.7 \pm 0.5$ for the natural-mix turbulence, and $N_{_{\rm leaf}} = 14.4 \pm 0.6$ and $N_{_{\rm level}} = 8.9 \pm 0.6$ for the purely compressive turbulence. Figure \ref{fig::coldendro} shows two example dendrograms built using the column density, one constructed from a natural mix turbulence simulation and the other from a purely compressive turbulence simulation. One can see that there is no significant difference between these dendrograms.

This means that dendrograms derived from column density maps are unable to accurately detect differences in the three-dimensional structure, and that one ought to be cautious when using them to compare different regions. Perhaps dendrograms constructed using PPV data may show the differences in structure, and this will be investigated in a future paper.

\subsection{The formation of fibre-like structures}

The formation of the fibre-like structures in situ shows that the fray and fragment scenario as proposed by \citet{TafHac15} is possible. A single main filament first forms due to the colliding flows. As mass is added to the main filament, it fragments to form elongated sub-structures, and eventually some of these fibre-like structures become unstable and fragment into cores. What is more common in these simulations is that hub networks appear and collapse occurs in the hubs. The appearance of hub networks is a small modification to the original scenario proposed by \citet{TafHac15}.

\begin{figure}
\centering
\includegraphics[width = 0.97\linewidth]{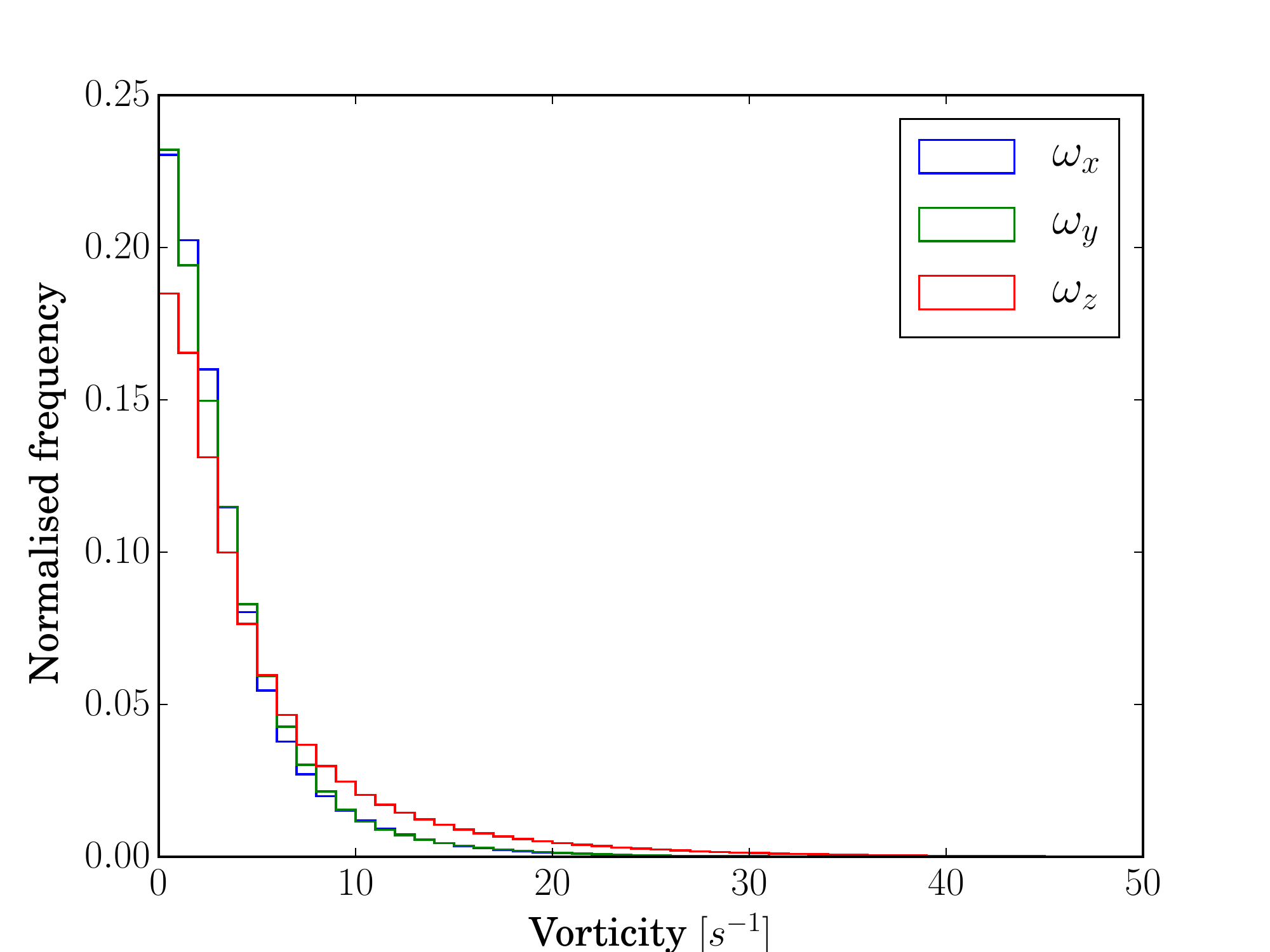}
\caption{The normalised distributions of the magnitude of the three vorticity components, $\omega_{x}$, $\omega_{y}$ and $\omega_{z}$, for an initially supersonic turbulent filament at $t=0.5 \, \rm Myr$. The distribution of the vorticity parallel to the longitudinal axis of the filament, $\omega_{z}$, shows a significant wing at large magnitudes.}
\label{fig::omegadis}
\end{figure}

Turbulence is the main driving force by which the filaments fragment into elongated sub-structures. Only the initially supersonic turbulence simulations produce such sub-structures, thus they require a high degree of turbulence to form. Gravity cannot be the dominant mechanism for the formation of these fibre-like structures, since gravitationally induced fragmentation causes collapse in the radial direction and produces the classic `beads on a string' cores. 

\citet{TafHac15} suggest that the fibres they observe are formed due to a combination of vorticity and self-gravity. Vorticity is defined as the curl of the velocity field, $\boldsymbol{\omega} \equiv \nabla \times \mathbf{v}$, and is a measure of the local spinning motion in a fluid.

Figure \ref{fig::omegadis} shows the normalised distributions of the magnitude of the three vorticity components, $\omega_{x}$, $\omega_{y}$ and $\omega_{z}$, for an initially supersonic turbulent filament at $t=0.5 \, \rm Myr$. One can see that $\omega_{x}$ and $\omega_{y}$ have near identical magnitude distributions. The vorticity component parallel to the longitudinal axis of the filament, $\omega_{z}$, has a similar distribution to the other components but with a large tail at high magnitude. This is likely a consequence of the filamentary geometry, accretion in the radial direction will primarily produce motions in the $x$-$y$ plane and appear as higher values of $\omega_{z}$.

\numberwithin{table}{section}

\begin{table}
\centering
\label{tab::spear}
\begin{tabular}{@{}*3l@{}}
\hline\hline
& \multicolumn{2}{c}{Spearman's $\rho$}\\ \hline
Velocity quantity & All densities & High densities \\
Divergence, $\nabla \cdot \textbf{v}$  & -0.16 & -0.11\\
Vorticity, $|\omega_x|$ & 0.31 & 0.28\\
Vorticity, $|\omega_y|$ & 0.32 & 0.30\\
Vorticity, $|\omega_z|$ & 0.52 & 0.47\\
Vorticity gradient, $|\delta \omega_z / \delta x|$ & 0.61 & 0.58\\ \hline
\end{tabular}
\centering
\caption{Results of Spearman's rank correlation tests for the correlation of different spatial derivatives of the velocity field with the volume-density.}
\end{table}

\begin{figure*}
\centering
\includegraphics[width = 0.47\linewidth]{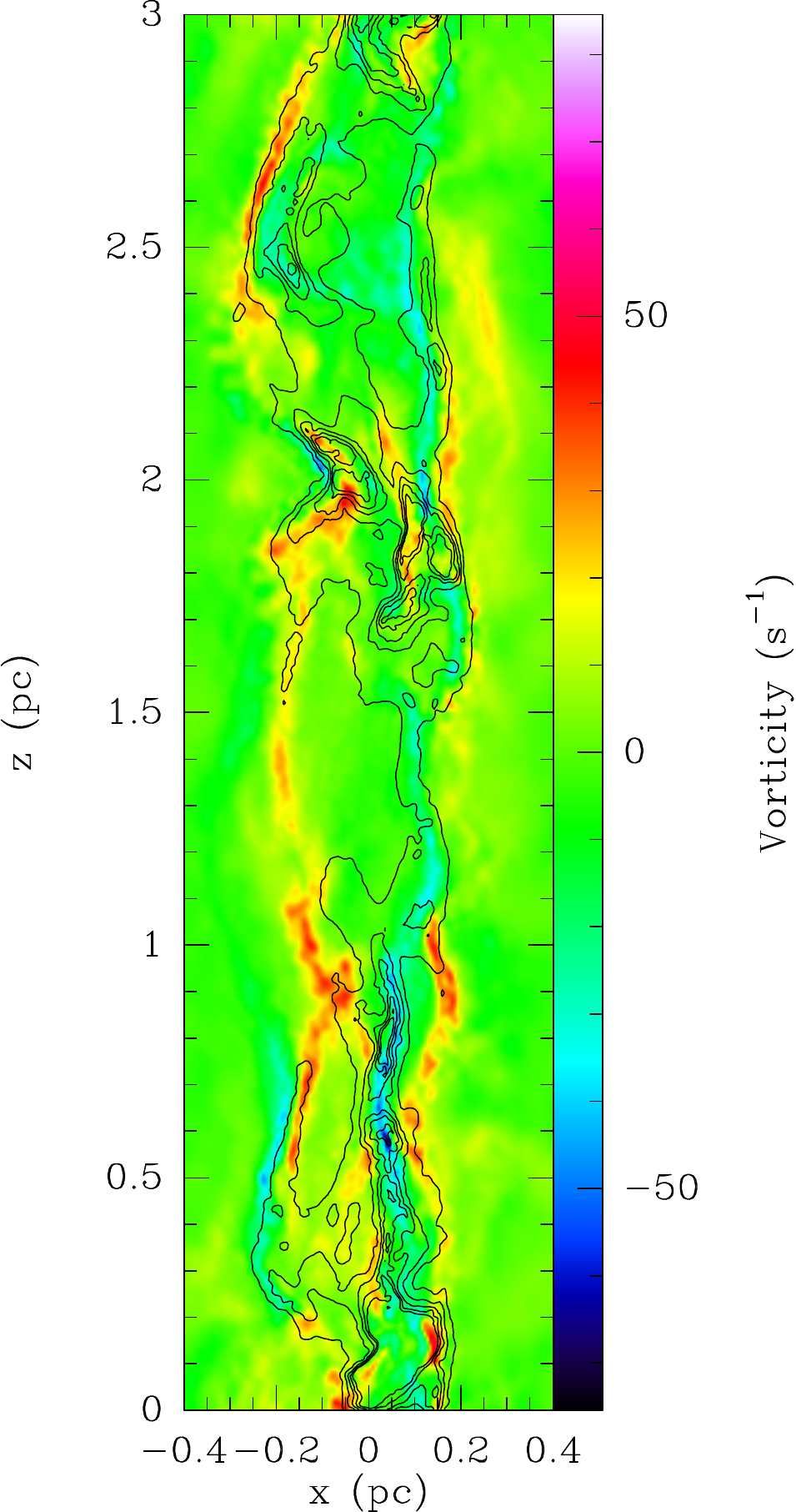}
\hspace{0.5cm}
\includegraphics[width = 0.47\linewidth]{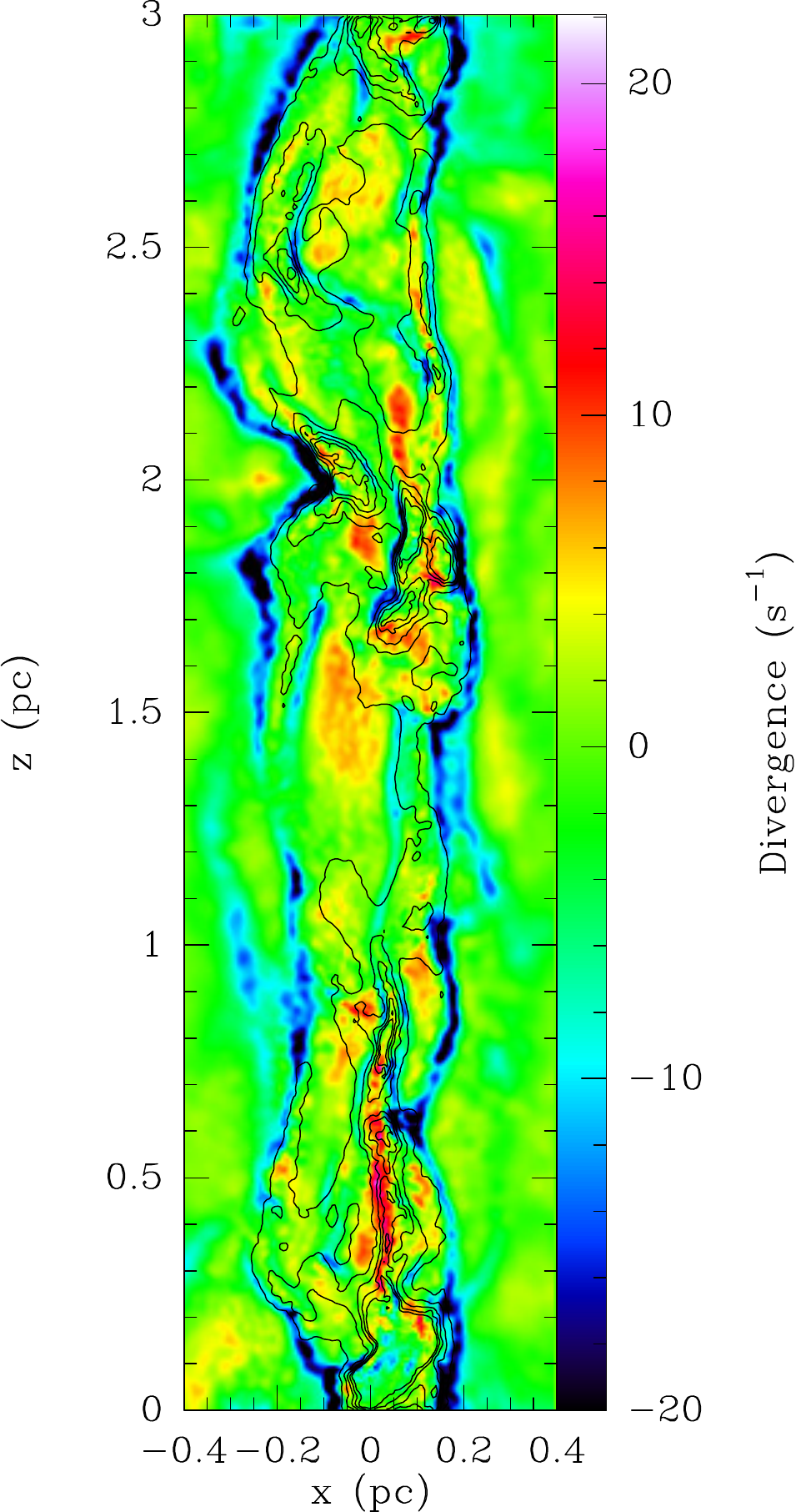}

\caption{(Left) A map of the vorticity component parallel to the longitudinal axis of the filament, the $z$-axis, at $y=0 \, \rm pc$ with density contours overlaid for an initially supersonic natural mix turbulence simulation. (Right) A map of the divergence of the velocity field in the filament, at $y=0 \, \rm pc$ with the same density contours. The density contours are at $2, 4, 6, 8, 10 \times 10^{-20} \, \rm{g/cm^{3}}$. One can see that the density structures follow the regions in which there exists a large vorticity gradient in the $x$-direction, i.e. where there are lines of blue and red/yellow next to each other. The divergence is seen to trace the accretion shock onto the main filament, but does not correlate well with the dense structures.}
\label{fig::vort}
\end{figure*}

\begin{figure*}
\centering
\includegraphics[width = 0.47\linewidth]{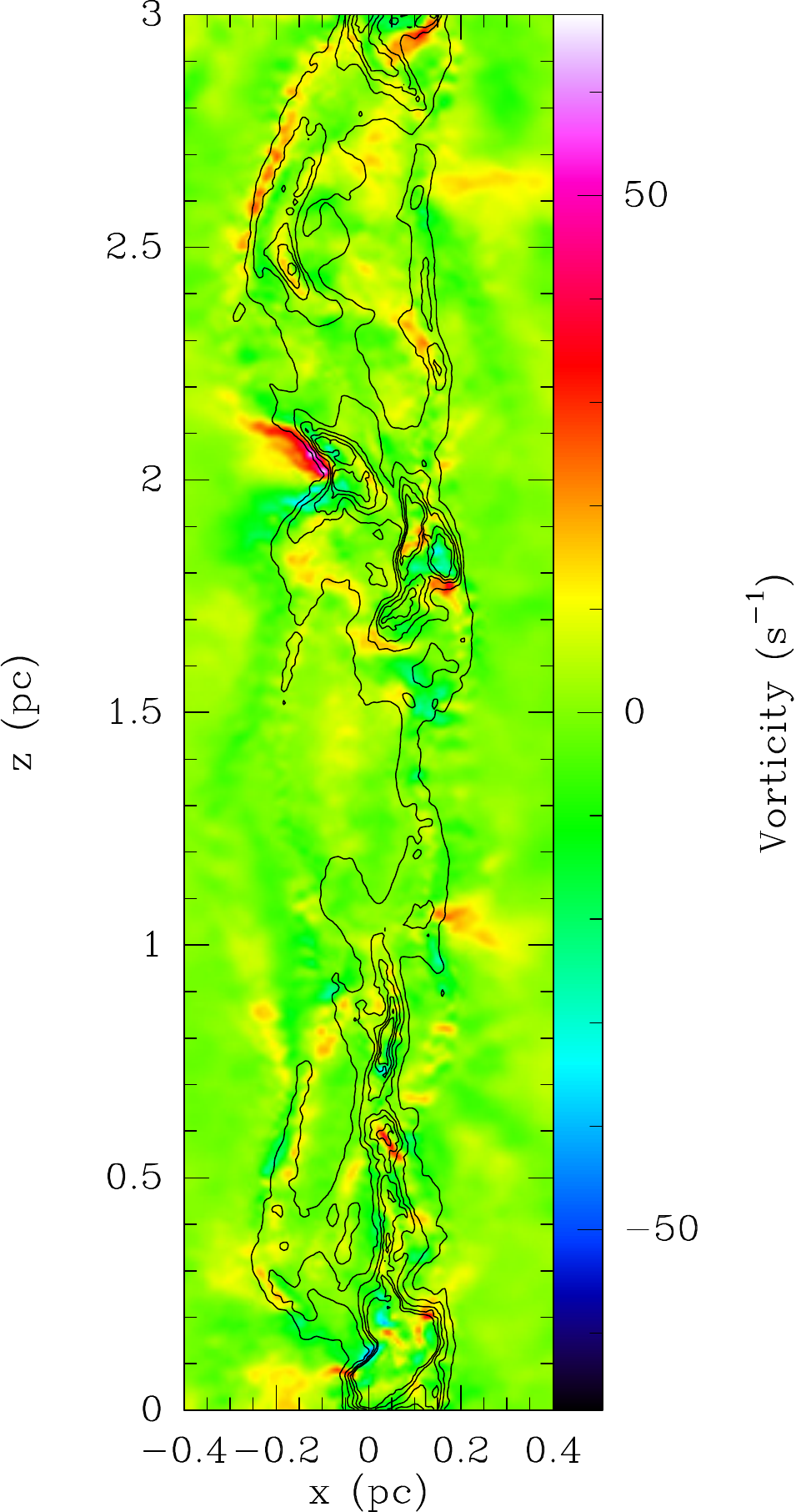}
\hspace{0.25cm}
\includegraphics[width = 0.47\linewidth]{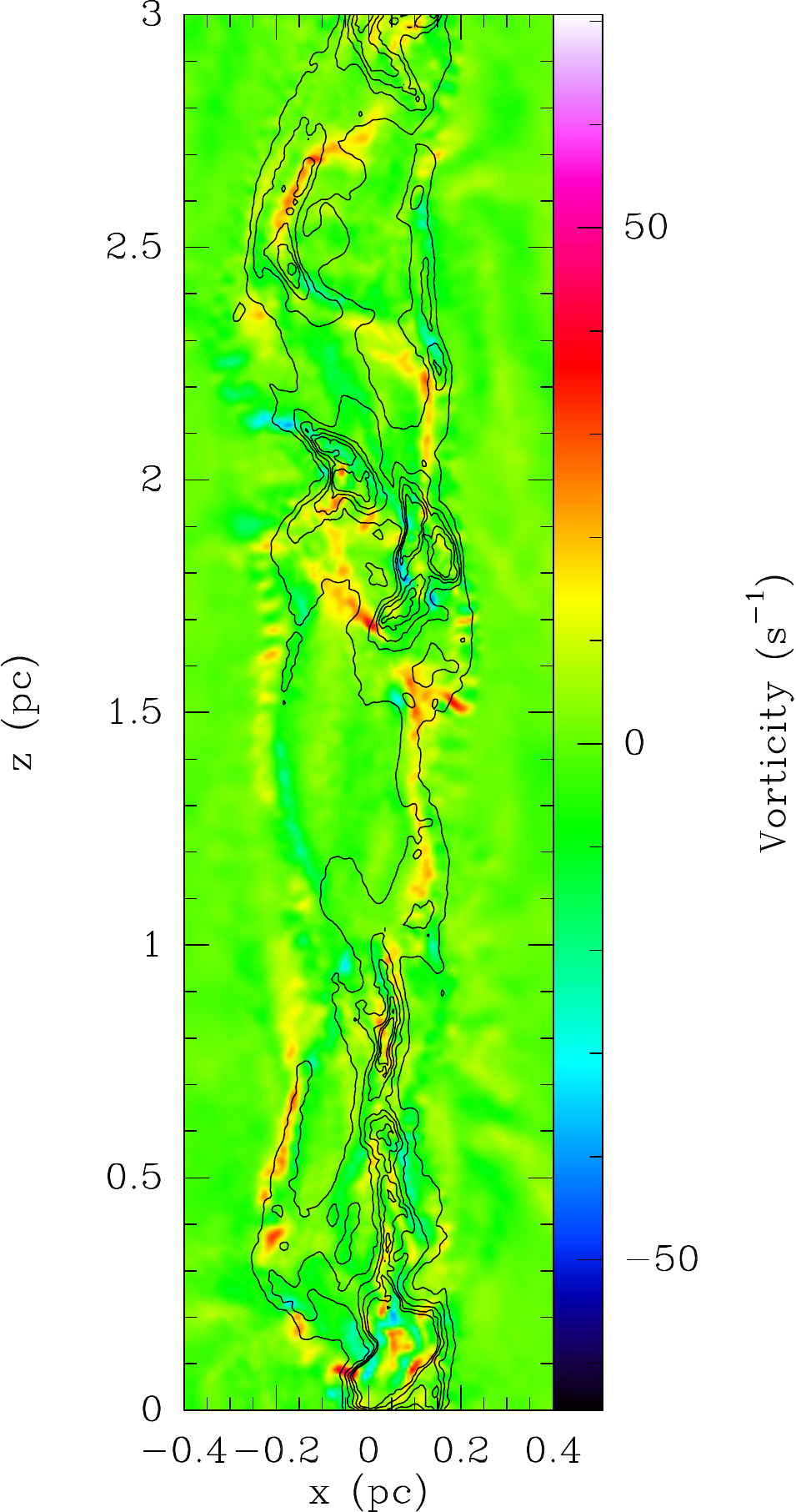}
\caption{A map of the vorticity component parallel to, on the left, the $x$-axis and, on the right, the component parallel to the $y$-axis at $y=0 \, \rm pc$ with density contours overlaid for an initially supersonic natural mix turbulence simulation. The density contours are at $2, 4, 6, 8, 10 \times 10^{-20} \, \rm{g/cm^{3}}$. Neither plot shows a strong correlation between vorticity and the density structure.}
\label{fig::wxwy}
\end{figure*}

Figure \ref{fig::vort} shows a map of the vorticity parallel to the filament's longitudinal axis, $\omega_z$, at $y = 0 \, \rm pc$, with volume density contours overlaid, for an initially supersonic turbulent filament at $t=0.5 \, \rm Myr$. The density contours are the same as those used in figure \ref{fig::los}. One can see that a correlation exists between vorticity and density; more precisely the dense elongated structures form in regions in which there exists a large vorticity gradient in the radial direction, $d \omega_z / dx$. Figure \ref{fig::vort} also shows a map of the divergence of the velocity field. The divergence traces the accretion shock onto the main filament but does not appear to correlate with the dense structures. There is also little correlation between the other two components of vorticity, $\omega_x$ and $\omega_y$, and density (figure \ref{fig::wxwy}).

To quantify the correlation between the properties of the velocity field and the density, we perform Spearman rank correlation tests. This gives a non-parametric measure of the correlation between two variables which are connected by a monotonic function. The results can be seen in table 1. As the direction of rotation is unimportant, the magnitude of the vorticity is used. The magnitude of the vorticity gradient is also used for the same reason. As the number of points considered is very large (80 x 80 x 300, the number of grid points), the corresponding $p-$values are vanishingly small, indicating the correlations are statistically significant. 

As one expects, the correlation between density and divergence is negative, i.e. dense material has a convergence flow while diffuse material has a divergent flow. However out of the five properties tested, divergence has the weakest correlation with density. Out of the three vorticity components, $|\omega_x|, |\omega_y|, |\omega_z|$, the vorticity component parallel to the filament's longitudinal axis, $|\omega_z|$, has the strongest correlation with density, as seen in figures \ref{fig::vort} and \ref{fig::wxwy}. Furthermore, the vorticity gradient, defined as $|\delta \omega_z / \delta x|$, shows an even stronger correlation with density, as can be seen in figure \ref{fig::vort}. When the sign of the vorticity and vorticity gradient is considered, Spearman's $\rho$ is roughly equal to zero, suggesting there is no directional preference. As can be seen in the third column of table 1, the same trends are apparent when one only considers high density regions, $\rho \, > \, 2 \, \times \, 10^{-20} \, \rm{g/cm}^{3}$ i.e. everything within the contours in figures \ref{fig::vort} and \ref{fig::wxwy}. The values of Spearman's $\rho$ are lower, likely due to smaller range over which the correlation is being measured.

 When one considers only the magnitude of the divergence, Spearman's $\rho = 0.30$; comparable to the correlations between $|\omega_x|$ and $|\omega_y|$, and density. This suggests that the positive correlation between $|\omega_x|$, $|\omega_y|$ and $|\nabla \cdot \textbf{v}|$, and density, is mainly due to the fact that the velocity field is more rapidly changing in higher density gas. The values of Spearman's $\rho$ are greater for $|\omega_z|$ and $|\delta \omega_z / \delta x|$, suggesting that the same argument is not capable of explaining the correlation between these quantities and density. It is possible that due to the larger amount of energy in $\omega_{z}$, as seen in figure \ref{fig::omegadis}, motions in this direction have a greater impact on structure formation. Thus the stronger correlation is the result of filamentary accretion primarily driving vorticity parallel to the longitudinal axis, leading to fibre-like structures forming.   

\begin{figure*}
\centering
\includegraphics[width = 0.47\linewidth]{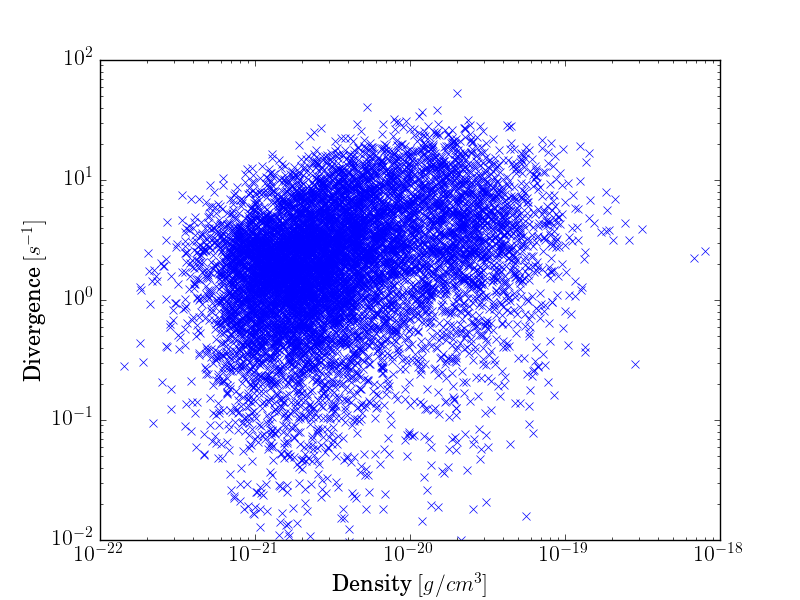}
\includegraphics[width = 0.47\linewidth]{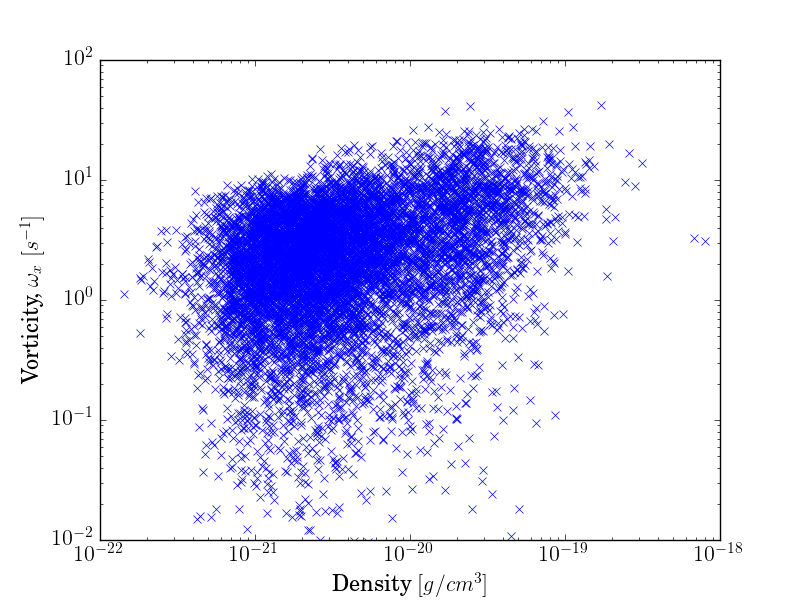}
\includegraphics[width = 0.47\linewidth]{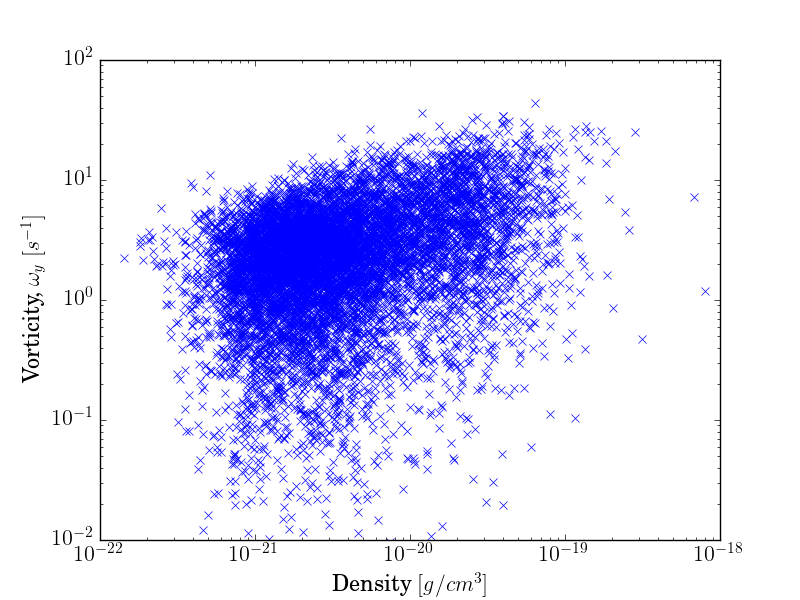}
\includegraphics[width = 0.47\linewidth]{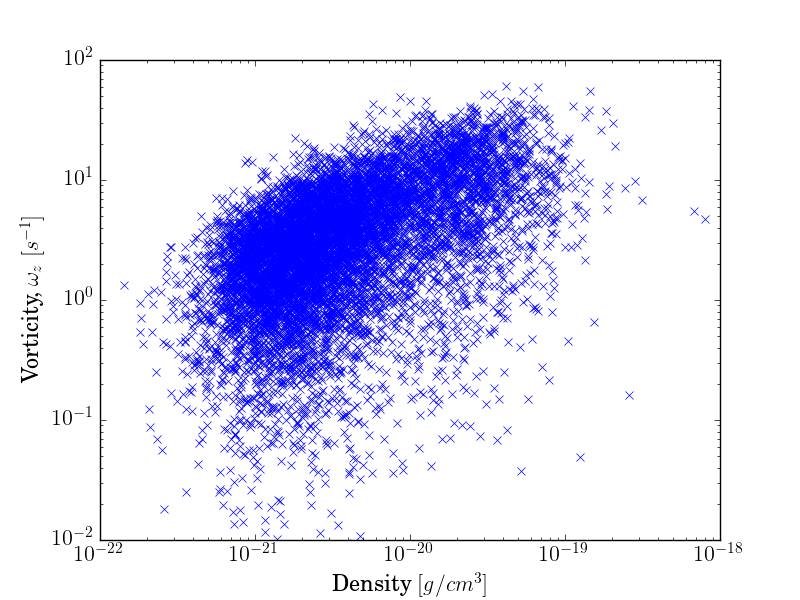}
\includegraphics[width = 0.47\linewidth]{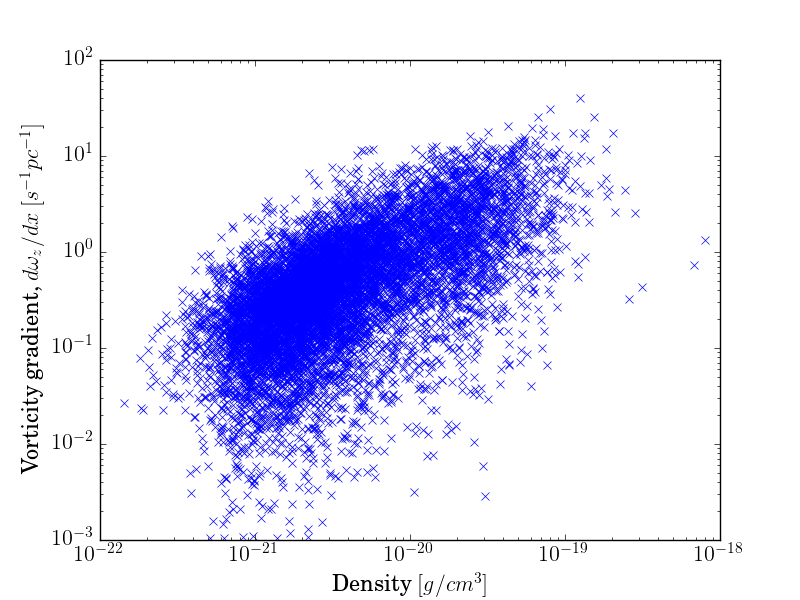}
\caption{Plots showing volume density against the velocity quantities listed in table \ref{tab::spear}. One can see that the relationship between $|\omega_{z}|$ and $|\delta \omega_z / \delta x|$, and density is both tighter and steeper than the link between the other velocity quantities and density.}
\label{fig::correlations}
\end{figure*}

Figure \ref{fig::correlations} shows density plotted against the velocity field properties in table 1. One can see that a correlation exists between all of these properties and density. Moreover, the relationship between $|\omega_{z}|$ and $|\delta \omega_z / \delta x|$, and density is both tighter and steeper than the link between the other velocity quantities and density; as expected from the results of the Spearman rank correlation tests. 

\begin{figure*}
\centering
\includegraphics[width = 0.9\linewidth,clip]{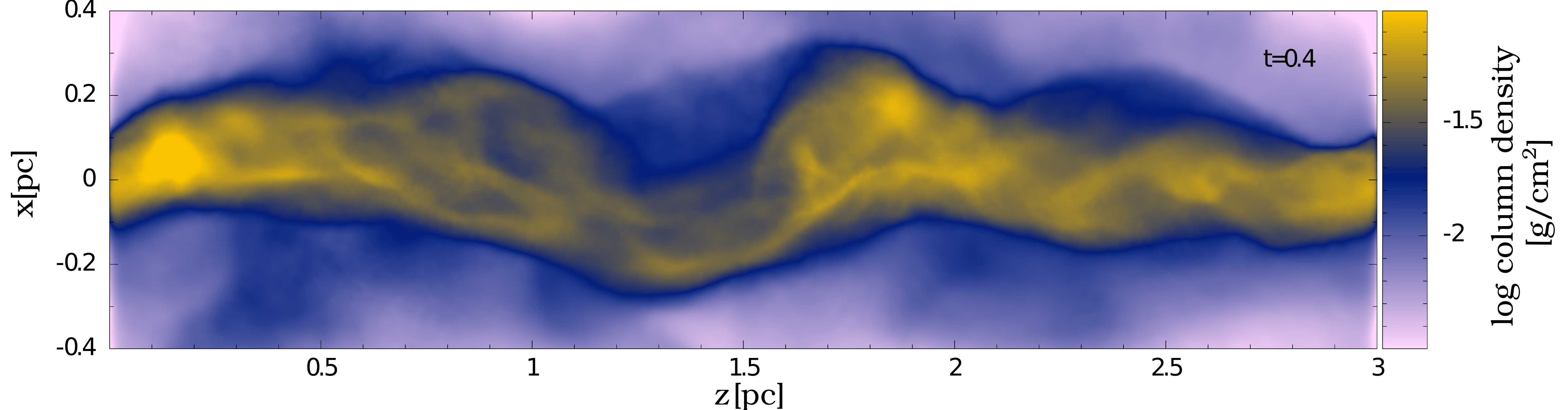}
\includegraphics[width = 0.9\linewidth,clip]{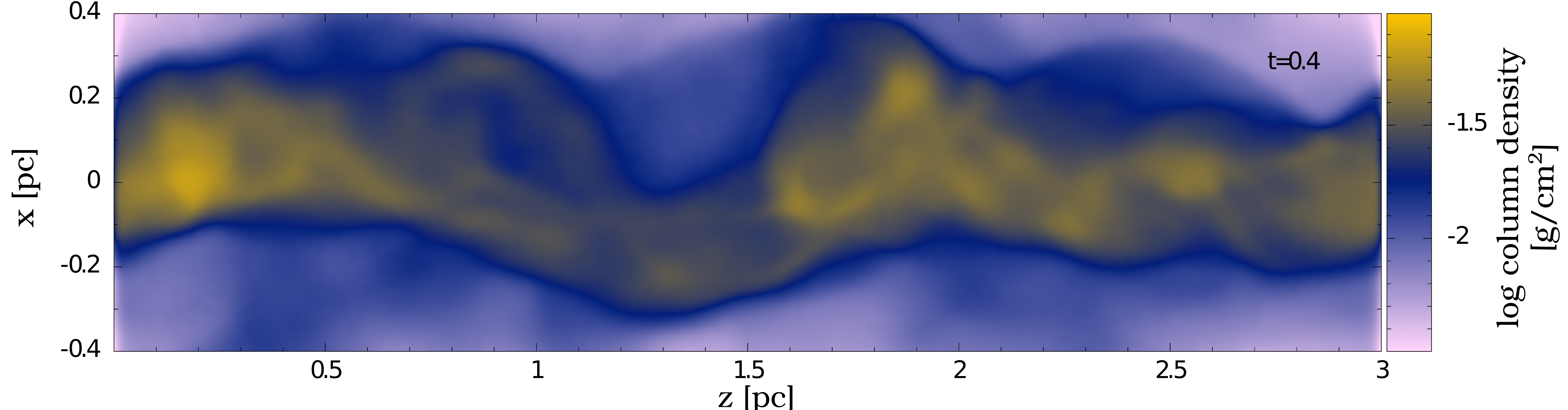}
\caption{Column density plots of two natural-mix, initially supersonic turbulence simulations which have the same initial conditions. The top panel shows the simulation with self-gravity, and the bottom panel shows the simulation without self-gravity. One can see that both simulations show the same elongated sub-structure, but the sub-structures are more well defined when self-gravity is included. Thus the velocity field is the dominant mechanism for the formation of these fibre-like structures, gravity acts against dispersal but is not necessary for their formation.}
\label{fig::gravnograv}
\end{figure*}

\begin{figure}
\centering
\includegraphics[width = 0.98\linewidth]{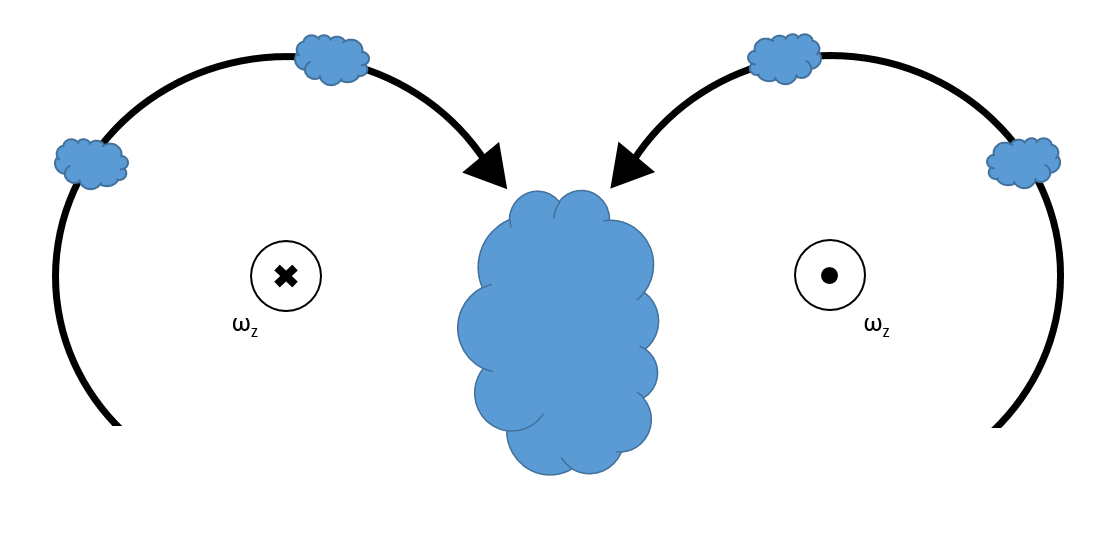}
\caption{A face down view illustrating how two nearby sites with anti-parallel vorticity vectors can produce regions of higher density. The two flows are able to bring gas into the centre from both direction where it can be held together by self-gravity and form a fibre-like structure.}
\label{fig::cartoon}
\end{figure}

Figure \ref{fig::cartoon} is a cartoon illustrating how two nearby sites with anti-parallel vorticity vectors will result in a build-up of material and produce sub-structure. The velocity field brings gas parcels in to the centre and forms an area of higher density, gravity can then act to stabilise this structure even though it did not form the structure. However, even with gravity acting against dispersion, as the fibre-like structures are typically sub-critical, they ought to be short-lived. This could explain the prevalence of `sterile' fibres found in Taurus \citep{Hac13,TafHac15}, which are predominately sub-critical; they are mainly transitory structures created by the filament's interior velocity field. Only the small proportion of fibres that acquire enough mass to become super-critical before they disperse are `fertile' and fragment to form cores. One would then expect an increase in the number of `fertile' fibres in a more massive and higher density filament, because the velocity field could produce more super-critical fibres before they disperse. 

To test that the turbulent velocity field is the dominant cause of the filament fragmenting into fibre-like structures, we re-run a simulation with no self-gravity. Figure \ref{fig::gravnograv} shows the column density plots of two simulations with the same initial conditions, but one with gravity and the other without. One can see that both simulations show the same general morphology and presence of elongated sub-structure. Gravity has therefore not played a dominant role in forming the fibre-like structures, but does act to hold them together, as seen in figure \ref{fig::gravnograv}a where the fibre-like structures are narrower and denser than in the simulation without gravity. This is consistent with the results presented in \citet{Fed16}, that turbulence is crucial for the structure of filaments but it is gravity that leads to higher peak column densities.

As the filament's internal turbulence tends to a roughly transonic level no matter what the initial level of turbulence is (as seen in figure \ref{fig::sig} and discussed in more detail in \S\ref{ssec::Acc}), the turbulence in the accretion flow must play a role in producing the differences in morphology between simulations with different initial levels of turbulence. Supersonic turbulence in the diffuse accretion flow produces density enhancements which flow onto the main filament. The accretion of this inhomogeneous flow onto the filament produces vorticity, evidenced by the high vorticity one can see at the boundary of the filament where the accretion shock is located (see figure \ref{fig::vort}). Thus the formation of fibre-like structures is intrinsically linked to a filament's accretion from a turbulent inhomogeneous medium. 

\subsection{Accretion driven turbulence}\label{ssec::Acc}

The convergence of the velocity dispersion in all cases to a roughly transonic level is interesting as observers have seen a number of such filaments \citep{Arz13,Hac13,Fer14,Kai16}. There must exist a continuous mechanism which produces and sustains turbulence in the filaments. \citet{Hei13a} propose that accretion flows onto a filament can drive turbulence. They conclude that the velocity dispersion driven by accretion is given by,
\begin{equation}
\sigma(t) = \left( 2 \epsilon R_f v^2(R_f) \frac{\dot{m}(t)}{m(t)} \right)^{1/3},
\label{eq::driven}
\end{equation} 
where $\epsilon$ is a driving efficiency factor in the range 0.01 - 0.1, $R_f$ is the filament's bounding radius, $v(R_f)$ is the velocity of the accreting gas at the filament's bounding radius, $m(t)$ is the filament's line-density at time $t$ and $\dot{m}(t)$ is the accretion rate at time $t$.

\begin{figure}
\centering
\includegraphics[width = 0.98\linewidth]{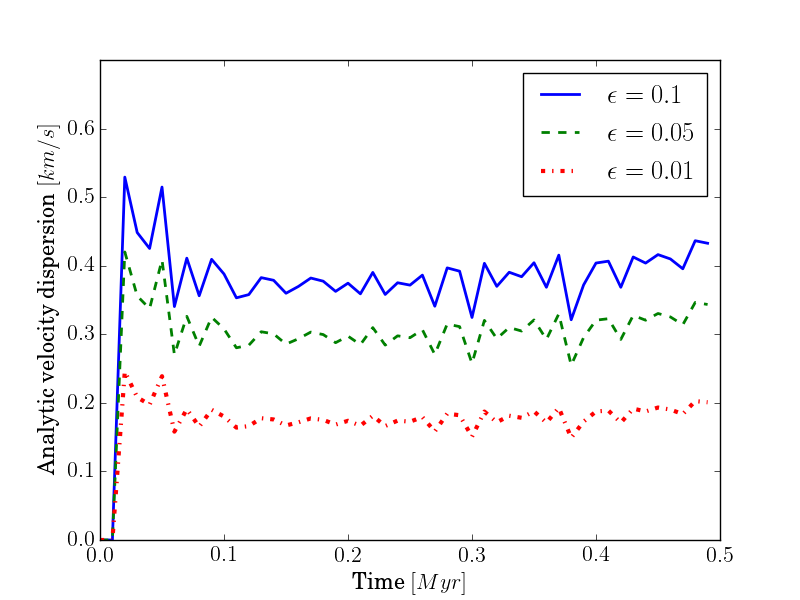}
\caption{Analytically estimated velocity dispersion due to driving by accretion for the subsonic natural mix turbulence simulations (equation \ref{eq::driven}). A driving efficiency between 5\% and 10\% is necessary to produce a velocity dispersion $\sim 0.4 \, \rm{km/s}$ as seen in the simulations.}
\label{fig::driven}
\end{figure}

One can use the information from the simulations to find the values of these parameters at every snapshot time and so produce a time evolution of the expected velocity dispersion due to driving. Figure \ref{fig::driven} shows the average expected velocity dispersion due to driving by accretion for $\epsilon=0.01, \, 0.05, \,  0.1$ using the 10 simulations with parameters $\delta_{\rm sol} \, = \, 2/3$ and $\sigma_{\rm 3D} \, = \, 0.1 \, \rm{km/s}$. An efficiency factor in the range of $5 - 10 \%$ is able to drive a velocity dispersion of $\sim 0.4 \, \rm{km/s}$; this is true for all 6 pairs of parameters.

\citet{SeiWal15} see a similar result in their simulations, the accretion is only able to drive a velocity dispersion of at most $0.5 \, \rm{km/s}$. The simulations presented in this chapter use a relatively high initial accretion rate onto the filament, $\dot{\mu} = 70 \, \rm{M_{\odot} \, pc^{-1} \, Myr^{-1}}$, which increases with time due to gravitational acceleration such that by the time of collapse $\dot{\mu} \sim 120  \, \rm{M_{\odot} \, pc^{-1} \, Myr^{-1}}$. Observations have shown that accretion rates on to filaments lie in the range $10 - 150 \, \rm{M_{\odot} \, pc^{-1} \, Myr^{-1}}$ \citep{Pal13,Kir13}, thus one cannot realistically invoke higher accretion rates to produce additional turbulence to stabilise nearby supercritical filaments against radial collapse. 

\begin{figure}
\centering
\includegraphics[width = 0.88\linewidth]{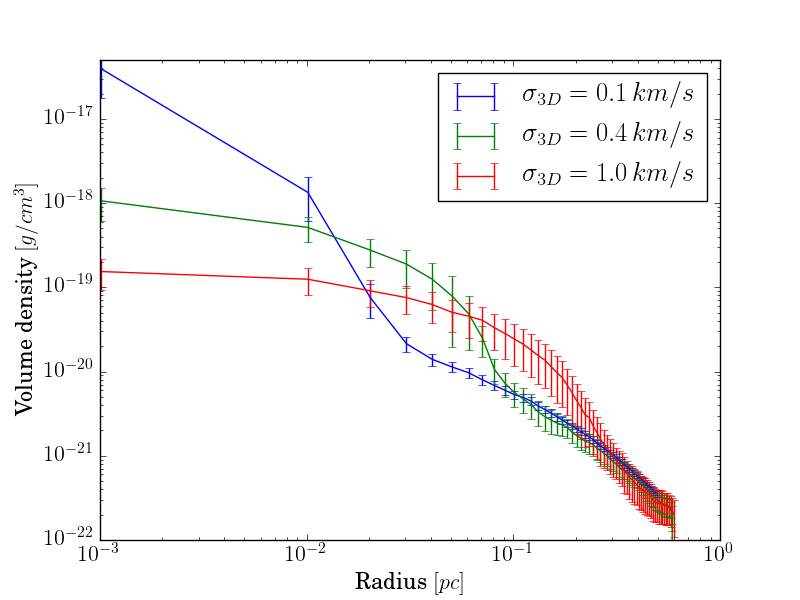}
\includegraphics[width = 0.88\linewidth]{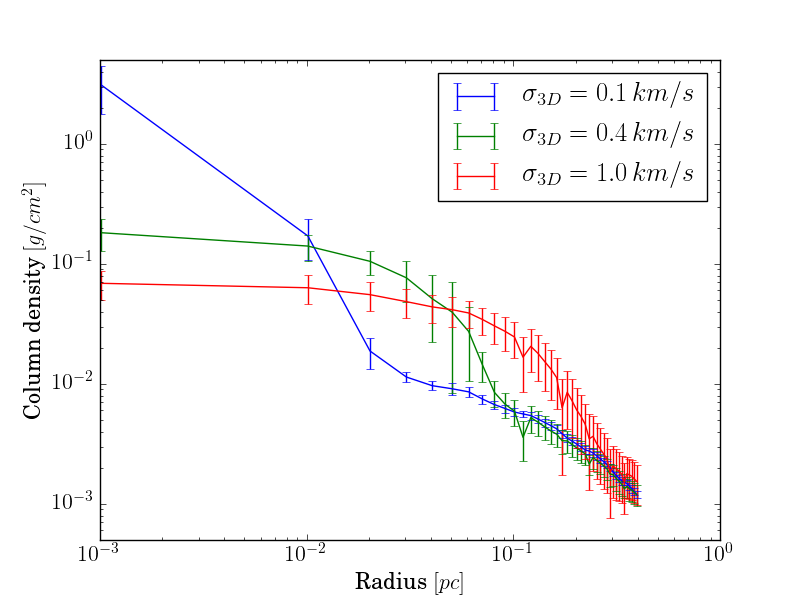}
\caption{(a) The azimuthally averaged radial volume density profiles for the three natural mix turbulent cases, initially sub-, tran- and supersonic velocity dispersions, at the end of the simulations when the filaments have line-densities $\sim 1.3 \, \mu_{_{\rm CRIT}}$. (b) The longitudinally averaged radial column density profiles for the same simulations. The error bars show the inter-quartile range around the median. }
\label{fig::denturb}
\end{figure}

However, while the initially sub- and transonic filaments do radially collapse and produce narrow filaments, the initially supersonic filaments are thermally super-critical at the end of the simulations but still show a large radial extent. Figure \ref{fig::denturb}a shows the azimuthally and longitudinally averaged radial volume density profiles for the three natural mix turbulent cases with initially sub-, tran- and supersonic velocity dispersions, at the end of the simulations when the filaments have line-densities $\sim 1.3 \, \mu_{_{\rm CRIT}}$. The median was used to average the profile so as to avoid undue influence of outliers, namely the very high density cores. One can see that the initially supersonic case shows the broadest radial profile. The full width half maxima (FWHMs) of these profiles are, $<0.02, \, <0.02$, and $0.08 \, \rm pc$ for the sub-, tran- and supersonic cases respectively. The sub- and transonic cases have upper limits as the resolution of the grid used to produce these plot is 0.01 pc. The same result is seen in column density (figure \ref{fig::denturb}b). The FWHMs of the column density profiles are, $<0.02, \, 0.02$, and $0.12 \, \rm pc$ for the sub-, tran- and supersonic cases respectively.

The fact that filaments growing from gas with supersonic turbulence do not collapse radially when their line-density exceeds the critical value can not be attributed to their internal velocity dispersion, since all three cases end up with roughly the same level of internal turbulence. The main difference between the three cases is the resulting morphology; it is possible that the considerable level of sub-structure in the supersonic cases helps to moderate the global radial instability. Possibly the break up of a filament into fibre-like structures invalidates the  assumptions underlying the derivation of the critical line-density in \citet{Ost64}; this needs to be investigated further, as it presents a possible solution to the problem that supercritical filaments should collapse radially to widths narrower than the observed $ 0.1 \, \rm pc$ \citep{Arz11,And16}.

\section{Conclusions}\label{SEC:CON}%

The process by which a filament fragments into cores is complicated by the presence of turbulence and inhomogeneous accretion. Moreover, the amount of turbulence and inhomogeneity produces significant differences in filament morphology and fragmentation. 

Filaments fragment gravitationally if they formed in a roughly sub-sonic turbulent and mildly inhomogeneous medium, or where gravity dominates over turbulence, and they show hierarchical two-tiered fragmentation, similar to the fragmentation seen in Orion \citep{Tei16,Kai16b}. The first fragments are large scale structures with wide separations, this fragmentation process is due to the gravitational fragmentation of the filament and is described by a modified dispersion relation derived in \citet{Cla16} (their equation 8). Next, these fragments undergo further fragmentation producing small scale separations, this fragmentation process occurs on the local Jeans length. This change in fragmentation behaviour on different length scales has been seen in G11 by \citet{Kai13}. Moreover, the fragmentation due to Jeans' instability occurs on a very short timescale, $\sim 0.1 \, \rm Myr$, and removes obvious signs of periodicity in the separation of fragments along a filament, possibly explaining why observed quasi-periodically fragmenting filaments are rare.

As the level of turbulence increases, gravity dominated filament fragmentation lessens in importance and the fragmentation length scales are mainly determined by the density perturbations delivered by the turbulence. As well as changing fragmentation length scales, increased turbulence causes elongated fibre-like structures to form within the main filament, which is reminiscent of the \textit{fray and fragment} scenario of fibre production in situ proposed by \citet{TafHac15}. We see that these structures are intimately linked to the velocity field in the filament. Where there exists a large gradient in vorticity, the velocity field is able to gather gas together and form dense structures. We propose that this vorticity is generated by the radial accretion of an inhomogeneous accretion flow. The radial accretion of gas generates vorticity primarily in the longitudinal direction, thus `fraying' the filament into numerous fibre-like structures.

One could expect that, as fibres accrete from the gas within the filament, which is less turbulent than the gas in the accretion flow, they might fragment in a manner similar to the filaments in the initially sub-sonic simulations. Fibres could therefore be expected to fragment in a quasi-periodic manner. However, the simulations presented here do not contain enough mass for this to occur.

Although the initial level of turbulence in these simulations differs by a factor of 10, the internal filament velocity dispersion always tends towards the sonic level and remains stable until collapse begins. We show that accretion driving is able to sustain this level of turbulence if one takes an efficiency factor of $5 - 10 \%$. One sees that despite the same level of internal velocity dispersion the filaments in the initially supersonic simulations do not collapse radially when their line-density exceeds the critical value, unlike the initially sub- and transonic cases which do collapse radially. It is possible that the considerable level of sub-structure present means that the fibre-like structures may collapse locally once they become super-critical, whilst the parent filament does not necessarily collapse globally.  

The formation, evolution and fragmentation of fibre-like structures will be studied in more detail in future work using simulations which include time-dependent chemistry with coupled heating and cooling \citep[as described in][]{GloCla12}. These simulations will include realistic thermodynamics, and will allow us to produce synthetic observations to determine if the fibre-like structures seen in PPP space are comparable to the fibres observers see in PPV space. As dendrograms constructed using column density are unable to adequately distinguish large differences in filament sub-structure in PPP space, these synthetic observations will also enable use to investigate the use of dendrograms in PPV space.      

\section{Acknowledgments}\label{SEC:ACK}%
SDC gratefully acknowledges the support of a STFC postgraduate studentship. APW and ADC gratefully acknowledge the support of a consolidated grant (ST/N000706/1) from the UK STFC. DAH acknowledges the support of the DFG cluster of excellence ``Origin and Structure of the Universe''. SDC would like to thank Paul Clark and Gary Fuller for their many comments which helped to improve this paper. This work was performed using the computational facilities of the Advanced Research Computing at Cardiff (ARCCA) Division, Cardiff University. 

\bibliographystyle{mn2e}
\bibliography{ref} 

\begin{appendices}

\section{Sink implementation}\label{APP:SINK}%

The standard sink creation criteria invoked in \citet{Hub13} are as follows:
\begin{itemize}
\item The candidate SPH particle must have a density greater than a pre-defined sink creation density, $\rho_{_{\textrm{SINK}}}$.
\item The candidate SPH particle must have the lowest gravitational potential of all of its neighbours, i.e. be the local gravitational potential minimum.
\item The candidate SPH particle's smoothing kernel must not overlap with existing sink particles's accretion radii. This prevents overlapping sink particle's forming.
\item The candidate SPH particle must have a density greater than $\rho_{_{\textrm{HILL}}}$, which is given by
\begin{equation}
\rho_{_{\textrm{HILL}}} = - \frac{3 (\underline{r}_{ij} \cdot \underline{a}_{ij})}{\pi G |\underline{r}_{ij}|^2 },
\end{equation}
where $\underline{r}_{ij}$ is the position vector between the candidate SPH particle $i$ and an already existing sink particle $j$, and $\underline{a}_{ij}$ is the acceleration vector between the two. This criterion allows an extended condensate which already has a sink particle in it to fragment only if the candidate SPH particle is a density peak that dominates the local gravitational field. 
\end{itemize}

These criteria are similar to those used in other implementations in both SPH \citep{Bat95, Jap05} and AMR \citep{Kru04,Fed10b}.

Since a new sink creation criterion has been introduced, we run a convergence test to show our choice of $\rho_{_{\textrm{SINK}}}$ is appropriate. Presented in table \ref{tab::sinkcon}.1 is: the time the first sink formed, its position along the filament and the number of sinks at the end of the simulation. One can see that above the sink creation $\rho_{_{\textrm{SINK}}} > 10^{-15} \, \rm{g \, cm^{-3}}$, the sinks are well converged.

\numberwithin{table}{section}

\begin{table}
\centering
\label{tab::sinkcon}
\begin{tabular}{llll}
\hline
$\rho_{_{\textrm{SINK}}}$ ($\rm{g \, cm^{-3}}$) & $t_{sink}$ ($\rm Myr $) & $z_{sink}$ ($ \rm pc$) & $N_{sink}$ \\ \hline \hline
$10^{-12}$        & 0.629 & 0.234       & 11 \\
$10^{-13}$        & 0.629 & 0.234       & 11 \\
$10^{-14}$        & 0.628 & 0.234       & 11 \\
$10^{-15}$        & 0.627 & 0.234       & 11 \\
$10^{-16}$        & 0.617 & 0.235       & 20 \\
$10^{-17}$        & 0.568 & 0.122/0.235 & 62 \\
\end{tabular}
\caption{A table showing the effect the sink creation density, $\rho_{_{\textrm{SINK}}}$, has on the time the first sink particle forms, $t_{sink}$, its position along the filament, $z_{sink}$, and the number of sinks at the end of simulation, $N_{sink}$. Note that the third column has two entries for a sink creation density of $10^{-17} \, \rm{g \, cm^{-3}}$, this is because two separate sinks form in the same snapshot.}
\end{table}  

\section{Resolution study}\label{APP:RES}%
\numberwithin{table}{section}

To study the effect that the numerical resolution of the simulation has on the fragmentation we varied the number of SPH particles for a simulation with $\sigma_{\rm 3D} = 0.1 \, \rm{km/s}$ and $\delta_{\rm sol} = 2/3$. To increase the effective resolution by a factor of 2, one needs to increase the number of particles by 8. Table \ref{tab::res}.1 shows the time the first sink formed, its position along the filament and the number of sinks at the end of the simulation. The number of sinks formed does not change as the resolution increases above 450,000 particles, and the position and time of the first sink changes by $<2\%$. Below 450,000 particles both the position of the first sink is different and the number of sinks formed increases. We are thus satisfied that these simulations are converged with 450,000 particles. 

\begin{table}
\centering
\label{tab::res}
\begin{tabular}{llll}
\hline
$N_{SPH}$ & $t_{sink}$ ($\rm Myr $) & $z_{sink}$ ($ \rm pc$) & $N_{sink}$ \\ \hline \hline
112,500        & 0.638 & 0.061       & 13 \\
225,000        & 0.621 & 0.062       & 16 \\
450,000        & 0.627 & 0.234       & 11 \\
900,000        & 0.633 & 0.234       & 11 \\
1,800,000      & 0.638 & 0.231       & 11 \\
3,600,000      & 0.640 & 0.233       & 11 \\

\end{tabular}
\caption{A table showing the effect the number of particles, $N_{SPH}$, has on the time the first sink particle forms, $t_{sink}$, its position along the filament, $z_{sink}$, and the number of sinks at the end of simulation, $N_{sink}$. One can see that it has very little effect once the resolution is above 450,000 particles.}
\end{table}

\section{Summary of sink properties}\label{APP:TAB}%
\numberwithin{table}{section}

Below are tables detailing the results of all 60 individual simulations. We present the time at which the first sink formed, $t_{sink}$, the time the simulation ended when $10\%$ of the mass is in the form of sinks, $t_{10\%}$, the difference between these times, $\Delta t$, and the number of sinks formed by the end of the simulation, $N_{sink}$, for simulations with initially subsonic (table \ref{tab::subsig}.1), transonic (table \ref{tab::transig}.2) and supersonic (table \ref{tab::supersig}.3) turbulence. 

\begin{table*}
\centering
\label{tab::subsig}
\begin{tabular}{lllllllll}
\hline
     & \multicolumn{4}{l}{Natural mix}                            & \multicolumn{4}{c}{Purely compressive}      \\ \hline
Seed & $t_{sink}$& $t_{10\%}$ & $\Delta t$  & $N_{sink}$ & $\quad$ $t_{sink}$ & $t_{10\%}$ & $\Delta t$ & $N_{sink}$ \\ \hline
1        & 0.627 & 0.697 & 0.070 & 11   & $\quad$ 0.594 & 0.676 & 0.082 & 8  \\
2        & 0.617 & 0.691 & 0.074 & 17   & $\quad$ 0.599 & 0.651 & 0.052 & 10 \\
3        & 0.564 & 0.632 & 0.068 & 10   & $\quad$ 0.579 & 0.650 & 0.071 & 10 \\
4        & 0.624 & 0.690 & 0.066 & 12   & $\quad$ 0.589 & 0.651 & 0.062 & 10 \\
5        & 0.620 & 0.681 & 0.061 & 11   & $\quad$ 0.590 & 0.658 & 0.068 & 9  \\
6        & 0.630 & 0.675 & 0.045 & 8    & $\quad$ 0.587 & 0.655 & 0.068 & 11 \\
7        & 0.621 & 0.693 & 0.072 & 13   & $\quad$ 0.620 & 0.661 & 0.041 & 14 \\
8        & 0.615 & 0.695 & 0.080 & 14   & $\quad$ 0.603 & 0.664 & 0.061 & 14 \\
9        & 0.651 & 0.697 & 0.046 & 11   & $\quad$ 0.603 & 0.659 & 0.056 & 14 \\
10       & 0.607 & 0.674 & 0.067 & 10   & $\quad$ 0.608 & 0.658 & 0.050 & 12 \\ \hline
Mean     & 0.618 & 0.683 & 0.065 & 11.7 & $\quad$ 0.597 & 0.658 & 0.061 & 11.2 \\
$\sigma$ & 0.021 & 0.019 & 0.011 & 2.4  & $\quad$ 0.011 & 0.007 & 0.011 & 2.1  
\end{tabular}
\caption{A table showing the time at which the first sink formed, $t_{sink}$, the time the simulation ended when $10\%$ of the mass is in the form of sinks, $t_{10\%}$, the difference between these times, $\Delta t$, and the number of sinks formed by the end of the simulation, $N_{sink}$, for each simulation with an initial sub-sonic velocity dispersion of $\sigma_{\rm 3D} = 0.1 \, \rm{km/s}$. Also shown is the mean and standard deviation of these statistics.}
\end{table*}

\begin{table*}
\centering
\label{tab::transig}
\begin{tabular}{lllllllll}
\hline
     & \multicolumn{4}{l}{Natural mix}                            & \multicolumn{4}{c}{Purely compressive}      \\ \hline
Seed & $t_{sink}$& $t_{10\%}$ & $\Delta t$  & $N_{sink}$ & $\quad$ $t_{sink}$ & $t_{10\%}$ & $\Delta t$ & $N_{sink}$ \\ \hline
1        & 0.606 & 0.748 & 0.142 & 12   & $\quad$ 0.557 & 0.689 & 0.132 & 8  \\
2        & 0.593 & 0.712 & 0.119 & 19   & $\quad$ 0.540 & 0.662 & 0.122 & 10 \\
3        & 0.472 & 0.776 & 0.304 & 4    & $\quad$ 0.514 & 0.634 & 0.120 & 3  \\
4        & 0.610 & 0.689 & 0.079 & 14   & $\quad$ 0.520 & 0.690 & 0.170 & 12 \\
5        & 0.646 & 0.721 & 0.075 & 9    & $\quad$ 0.508 & 0.633 & 0.125 & 15 \\
6        & 0.610 & 0.724 & 0.114 & 10   & $\quad$ 0.570 & 0.689 & 0.119 & 10 \\
7        & 0.641 & 0.734 & 0.093 & 10   & $\quad$ 0.655 & 0.736 & 0.081 & 13 \\
8        & 0.607 & 0.704 & 0.097 & 13   & $\quad$ 0.569 & 0.705 & 0.136 & 11 \\
9        & 0.702 & 0.776 & 0.074 & 12   & $\quad$ 0.528 & 0.642 & 0.114 & 10 \\
10       & 0.588 & 0.687 & 0.099 & 16   & $\quad$ 0.611 & 0.689 & 0.078 & 12 \\ \hline
Mean     & 0.608 & 0.727 & 0.120 & 11.9 & $\quad$ 0.557 & 0.678 & 0.120 & 10.4 \\
$\sigma$ & 0.055 & 0.030 & 0.065 & 3.9  & $\quad$ 0.044 & 0.032 & 0.025 & 3.1  
\end{tabular}
\caption{A table showing the time at which the first sink formed, $t_{sink}$, the time the simulation ended when $10\%$ of the mass is in the form of sinks, $t_{10\%}$, the difference between these times, $\Delta t$, and the number of sinks formed by the end of the simulation, $N_{sink}$, for each simulation with an initial sonic velocity dispersion of $\sigma_{\rm 3D} = 0.4 \, \rm{km/s}$. Also shown is the mean and standard deviation of these statistics.}
\end{table*}

\begin{table*}
\centering
\label{tab::supersig}
\begin{tabular}{lllllllll}
\hline
     & \multicolumn{4}{l}{Natural mix}                            & \multicolumn{4}{c}{Purely compressive}      \\ \hline
Seed & $t_{sink}$& $t_{10\%}$ & $\Delta t$  & $N_{sink}$ & $\quad$ $t_{sink}$ & $t_{10\%}$ & $\Delta t$ & $N_{sink}$ \\ \hline
1        & 0.652 & 0.864 & 0.212 & 13   & $\quad$ 0.684 & 0.811 & 0.127 & 9  \\
2        & 0.665 & 0.793 & 0.128 & 20   & $\quad$ 0.661 & 0.735 & 0.084 & 12 \\
3        & 0.436 & 0.658 & 0.222 & 7    & $\quad$ 0.469 & 0.585 & 0.116 & 11 \\
4        & 0.695 & 0.859 & 0.164 & 13   & $\quad$ 0.532 & 0.718 & 0.186 & 13 \\
5        & 0.911 & 1.027 & 0.116 & 23   & $\quad$ 0.565 & 0.693 & 0.128 & 12 \\
6        & 0.801 & 0.867 & 0.066 & 11   & $\quad$ 0.453 & 0.641 & 0.188 & 9  \\
7        & 0.730 & 0.847 & 0.117 & 16   & $\quad$ 0.836 & 0.892 & 0.056 & 9  \\
8        & 0.684 & 0.820 & 0.136 & 7    & $\quad$ 0.688 & 0.814 & 0.126 & 18 \\
9        & 0.930 & 1.010 & 0.080 & 3    & $\quad$ 0.516 & 0.659 & 0.143 & 13 \\
10       & 0.661 & 0.725 & 0.064 & 14   & $\quad$ 0.701 & 0.837 & 0.136 & 10 \\ \hline
Mean     & 0.717 & 0.847 & 0.131 & 12.7 & $\quad$ 0.611 & 0.739 & 0.128 & 11.6 \\
$\sigma$ & 0.134 & 0.107 & 0.052 & 5.8  & $\quad$ 0.116 & 0.093 & 0.039 & 2.6  
\end{tabular}
\caption{A table showing the time at which the first sink formed, $t_{sink}$, the time the simulation ended when $10\%$ of the mass is in the form of sinks, $t_{10\%}$, the difference between these times, $\Delta t$, and the number of sinks formed by the end of the simulation, $N_{sink}$, for each simulation with an initial supersonic velocity dispersion of $\sigma_{\rm 3D} = 1.0 \, \rm{km/s}$. Also shown is the mean and standard deviation of these statistics.}
\end{table*}

\end{appendices}

\label{lastpage}

\end{document}